\documentclass[10pt, twocolumn, comsoc]{IEEEtran}

\usepackage{graphicx,epsfig}
\usepackage[noadjust]{cite}
\usepackage{mcite}
\usepackage{amsfonts,helvet}
\usepackage{fancyhdr}
\usepackage{threeparttable}
\usepackage{epsf,epsfig}
\usepackage{amsthm}
\usepackage{amsmath}
\usepackage{siunitx}
\usepackage{amssymb}
\usepackage{dsfont}
\usepackage{subfigure}
\usepackage{color}
\usepackage[linesnumbered,ruled]{algorithm2e}
\usepackage{enumerate}
\usepackage{cancel}
\usepackage{comment}
\usepackage{booktabs}
\usepackage{multirow}

\newtheorem{lemma}{Lemma}

\newtheorem{remark}{Remark}

\SetKwInput{KwInput}{Initialize}
\SetKwInput{KwOutput}{Output}
\SetKwProg{Fn}{Stage 1}{}{}
\SetKwProg{Fn}{Stage 2}{}{}
\SetKwSwitch{Switch}{Case}{Other}{Switch}{:}{Case}{Other}{EndCase}{EndSwitch}

\newcommand*{\myproofname}{Proof}
\newenvironment{myproof}[1][\myproofname]{\begin{proof}[#1]}{\end{proof}}

\usepackage{eucal}

\begin{document}

\title{ Distributed Precoding for Satellite-Terrestrial Integrated Networks Without Sharing CSIT: \\ A Rate-Splitting Approach }

\author{Doseon~Kim, Sungyoon~Cho, Wonjae~Shin, Jeonghun~Park, and Dong~Ku~Kim

\thanks{

Doseon Kim, Dong Ku Kim, and Jeonghun Park are with the School of Electrical and Electronic Engineering, Yonsei University, Seoul 03722, South Korea (e-mail: {\texttt{kds1018@yonsei.ac.kr; jhpark@yonsei.ac.kr; dkkim@yonsei.ac.kr}}).
Sungyoon Cho is with the Korea Electronics Technology Institute, Seoul 03924, South Korea (email: {\texttt{sycho@keti.re.kr}}),
Wonjae Shin is with the School of Electrical Engineering, Korea University, Seoul 02841, South Korea (email: {\texttt{wjshin@korea.ac.kr}})
}}

\maketitle \setcounter{page}{1} 

\begin{abstract} 
Satellite-terrestrial integrated networks (STINs) are promising architecture for providing global coverage. 
In STINs, full frequency reuse between a satellite and a terrestrial base station (BS) is encouraged for aggressive spectrum reuse, which induces non-negligible amount of interference.
To address the interference management problem in STINs, this paper proposes a novel distributed precoding method. 
Key features of our method are: \lowercase\expandafter{\romannumeral1}) a rate-splitting (RS) strategy is incorporated for efficient interference management and \lowercase\expandafter{\romannumeral2}) the precoders are designed in a distributed way without sharing channel state information between a satellite and a terrestrial BS. 
Specifically, to design the precoders in a distributed fashion, we put forth a spectral efficiency decoupling technique, that disentangles the total spectral efficiency function into two distinct terms, each of which is dependent solely on the satellite's precoder and the terrestrial BS's precoder, respectively.
Then, to resolve the non-smoothness raised by the RS strategy, we approximate the spectral efficiency expression as a smooth function by using the LogSumExp technique; thereafter we develop a generalized power iteration inspired optimization algorithm built based on the first-order optimality condition. 
Simulation results demonstrate that the proposed method offers considerable spectral efficiency gains compared to the existing methods. 
\end{abstract}


\section{Introduction}

Managing inter-cell interference (ICI) is one of long-standing problems in cellular networks. Due to an inherent characteristic of resource reuse, ICI fundamentally limits the rate performance \cite{lozano:tit:13}; thus efficiently handling the interference is a key to achieve the high spectral efficiency of cellular networks. So far, numerous studies have been conducted to mitigate ICI by using multi-cell multiple-input multiple-output (MIMO) cooperative transmission \cite{gesbert:jsac:10}. 
In most of the MIMO cooperation techniques, a key principle is designing precoders in a coordinated fashion (i.e., coordinated precoding), where multiple base stations (BSs) share channel state information at transmitters (CSIT), then obtain their precoders by jointly exploiting the shared CSIT. 

Recently, with the growing interest in satellite communications, satellite-terrestrial integrated networks (STINs) gain significant attention. One promising scenario of operating STINs is using satellites to serve alienated users left outside of a terrestrial coverage region \cite{park:arxiv:23}. 
In such a scenario, assuming full spectrum sharing between a satellite and a terrestrial BS as considered in international telecommunication union (ITU) world radiocommunication conferences (WRC) 27 and \cite{Vazquez:wcom:16}, the interference stemming from a satellite is considered as a significant factor to deteriorate the sum spectral efficiency of STINs \cite{yin:twc:23, lee:tvt:24}. 
To handle this interference, one might be tempted to apply the widely used coordinated precoding approaches from previous researches, requiring to share CSIT between a satellite and a terrestrial BS.
Nonetheless, CSIT sharing between a satellite and a terrestrial BS is difficult. Unlike terrestrial BSs connected via wireline transport network of X2, satellites require extra wireless resources for connecting to a terrestrial gateway \cite{yin:twc:23}. 
For this reason, sharing CSIT in STINs causes significant overheads, which hinders the use of the conventional coordinated precoding methods. 
Accordingly, a desirable way to address the interference in STINs is using a distributed precoding method, where each of satellite and terrestrial BS determines their precoders individually while not sharing CSIT. Motivated by this, we propose a novel distributed precoding method for efficiently managing the interference in STINs.

\subsection{Related Work}
In design of distributed downlink precoding, it is infeasible to compute the exact signal-to-interference-plus-noise ratio (SINR) because CSIT sharing is not allowed. 
For this reason, it is critical to identify an alternative metric to take the place of SINR. 
For example, in \cite{sadek:twc:07}, the signal-to-leakage-plus-noise ratio (SLNR) was introduced, which is computed by the ratio between the signal power and the sum of the leakage interference plus the noise power. Here, the leakage interference power indicates the amount of transmit power that a BS incurs to neighboring cells. By doing this, one can readily come up with downlink precoding vectors by taking advantage of equivalence between the SLNR and the uplink SINR, wherein the optimal combiner is well-known as a minimum mean-squared error (MMSE). 
Using SLNR, in \cite{tran:twc:18}, transmission methods for spectral efficiency and energy efficiency were proposed. 
Extending SLNR to multi-point joint transmission setups, in \cite{zakhour:wsa:09, zakhour:twc:10, bjornson:tsp:10}, virtual SINR (VSINR) was studied. Especially, in \cite{bjornson:tsp:10}, the Pareto optimal boundary of linear precoding was characterized using VSINR.

Even though SLNR is a tractable metric for distributed precoding, the performance degradation is unavoidable since it fails to appropriately account for the impact of inter-user interference (IUI). 
To overcome this, in \cite{Choi:twc:12}, the IUI term and the ICI term are detached from the sum spectral efficiency by exploiting high signal-to-noise ratio (SNR) approximation. Then the distributed precoding scheme was developed for jointly mitigating the IUI and the leakage.
In \cite{han:tcom:21}, a new metric called signal-to-interference-plus-leakage-plus-noise ratio (SILNR) was proposed, which primarily differs from SLNR in how it handles the IUI. 
In \cite{he:jsac:14}, assuming heterogeneous network scenarios, a per-cell energy efficiency maximization precoding was developed based on weighted MMSE (WMMSE), while restricting the maximum amount of leakage to other cells.
In \cite{wang:twc:22}, a distributed precoding method with network virtualization was proposed in a semi-closed form.

Despite the abundant prior work of distributed precoding methods, the existing schemes are inherently limited in an information-theoretical sense. To be specific, the existing methods implicitly assume a treating interference as noise (TIN) based decoding strategy, wherein each user only decodes its own signal without using an advanced treatment to interference. 
As shown in \cite{etkin:tit:08}, TIN is optimal in a weak interference regime, which indeed achieves quasi-optimal spectral efficiency \cite{bazco:wclett:19}. 
In a medium or a strong interference regime, however, TIN loses its optimality, rather it is far from the optimum \cite{etkin:tit:08}. 
Unfortunately, the considered STINs typically do not operate in a weak interference regime. 
This is because it is challenging to perfectly estimate CSIT in the STINs due to high randomness associated with satellite links \cite{You:jsac:20}. 
Additionally, to achieve high throughput in satellite communications, beamforming is used to serve multiple users simultaneously \cite{perez:sigmag:19}. 
In this case, since only imperfect CSIT is available, significant amount of interference can be caused, where the interference power becomes even more strong by considering the low altitude of low-earth-orbit (LEO) satellites. This leads the STINs to a medium-strong interference regime.
In such a regime, a proactive decoding strategy such as rate-splitting (RS) is more appealing \cite{park:network:23}.

There exists some prior work that employed RS in STINs based on satellite-terrestrial BS coordination. 
For instance, in \cite{yin:twc:23}, two approaches were proposed: 
one is a coordinated approach that shares CSIT, the other is a cooperative approach that shares not only CSIT but also transmitted data.
For both approaches, precoding schemes incorporating RS were proposed to mitigate the interference in STINs. 
In \cite{lee:tvt:24}, a coordinated approach of \cite{yin:twc:23} was extended incorporating multi-layer RS.
In \cite{Yunnuo:arXiv:23}, a multi-layer interference management scheme was proposed in multiple satellite networks, where the RS strategy is implemented across different satellites through CSIT sharing. 
In \cite{Li:jsac:20, Zhao:twc:23, park:twc:23, joudeh:tcom:16}, even though they did not particularly consider a STIN scenario, novel precoding optimization methods was proposed for RS setups.
For such methods \cite{yin:twc:23, lee:tvt:24, Yunnuo:arXiv:23}, CSIT or transmitted data should be shared between a satellite and a terrestrial BS. As mentioned above, this sharing is not straightforward in STINs due to the absence of a dedicated link between a satellite and a terrestrial BS. 
As a result, there is a need to develop a precoding method that \lowercase\expandafter{\romannumeral1}) incorporates RS and \lowercase\expandafter{\romannumeral2}) requires no CSIT sharing. 
This serves as a core motivation of our work.

\subsection{Contributions}
In this paper, we consider a downlink STIN, where a terrestrial BS serves terrestrial users (TUs) located within its coverage, while a LEO satellite serves satellite users (SUs) situated beyond the coverage range of the terrestrial BS.
For this reason, the terrestrial BS does not impose any interference to the SUs. 
On the contrary, the satellite can incur the interference to certain TUs in a region reachable by the LEO satellite signal \cite{yin:twc:23, lee:tvt:24}.
Given this setup, we devise a distributed precoding method to efficiently handle the interference without sharing CSIT. 
Key features of our method are summarized as follows. 

\begin{itemize}
    \item {\textbf{RS strategy}}: 
    We employ the RS strategy, which enables the SUs and a part of the TUs to decode the interference coming from the satellite by using successive interference cancellation (SIC). 
    To this end, the messages intended to the SUs are split into a common and a private parts, wherein the common parts are jointly encoded into a common stream decodable by the SUs and a part of the TUs. 
    In decoding, the SUs and a part of the TUs decode and eliminate the common stream while treating other streams as noise, and then decode the private stream. Upon this RS decoding process, we derive a lower bound on the spectral efficiencies by considering the effects on the imperfect CSIT and the RS decoding condition. Subsequently, we formulate the sum spectral efficiency maximization problem.

    \item {\textbf{Spectral efficiency decoupling}}:
    Addressing the posed problem necessitates a coordinated approach involving CSIT sharing between the LEO satellite and the terrestrial BS.
    To solve this problem in a distributed fashion, we present a novel spectral efficiency decoupling technique.
    Specifically, we take average on the IUI terms over the randomness related to incomplete knowledge of the channel fading process and precoder design. Then we decouple the spectral efficiencies into two isolated terms, each of which is only associated with a terrestrial BS precoding vector and a satellite precoding vector, respectively. Based on this, we transform the original problem into the distributed precoding optimization problem.

    \item {\textbf{Distributed precoding optimization}}: 
    Even after transforming the original problem into the distributed problem, finding its solution is challenging since the problem is non-convex and non-smooth. To address this, we first approximate the objective function by using the LogSumExp (LSE) technique. 
    Then we characterize the first-order Karush–Kuhn–Tucker (KKT) conditions and cast these as a nonlinear eigenvalue problem with eigenvector dependency (NEPv). 
    We show that finding the principal eigenvector of these nonlinear eigenvalue problems is equivalent to finding the local optimal solution. 
    Leveraging this, we propose an iterative algorithm named STIN-generalized power iteration (STIN-GPI).

\end{itemize}

Through simulations, we numerically demonstrate that the proposed STIN-GPI algorithm outperforms the current state-of-the-art precoding methods in terms of the spectral efficiency and also the computational complexity. 

{\color{black}{
The paper is organized as follows. Section \uppercase\expandafter{\romannumeral2} introduces the system models considered in the paper, including the network model, the channel model, the signal model, and the performance metrics of the considered STIN. We also explain our main problem in Section \uppercase\expandafter{\romannumeral2}. 
In what follows, we transform the problem in a distributed manner to devise a distributed precoding method in Section \uppercase\expandafter{\romannumeral3}. 
By leveraging this, we propose our main method in Section \uppercase\expandafter{\romannumeral4} and present numerical results in Section \uppercase\expandafter{\romannumeral5}, respectively. Section \uppercase\expandafter{\romannumeral6} concludes the paper. 
}}


\section{System Models}

\subsection{Network Model}
\begin{figure}[t]
 \renewcommand{\figurename}{Fig.}
    \centering
    \includegraphics[width=9.5cm]{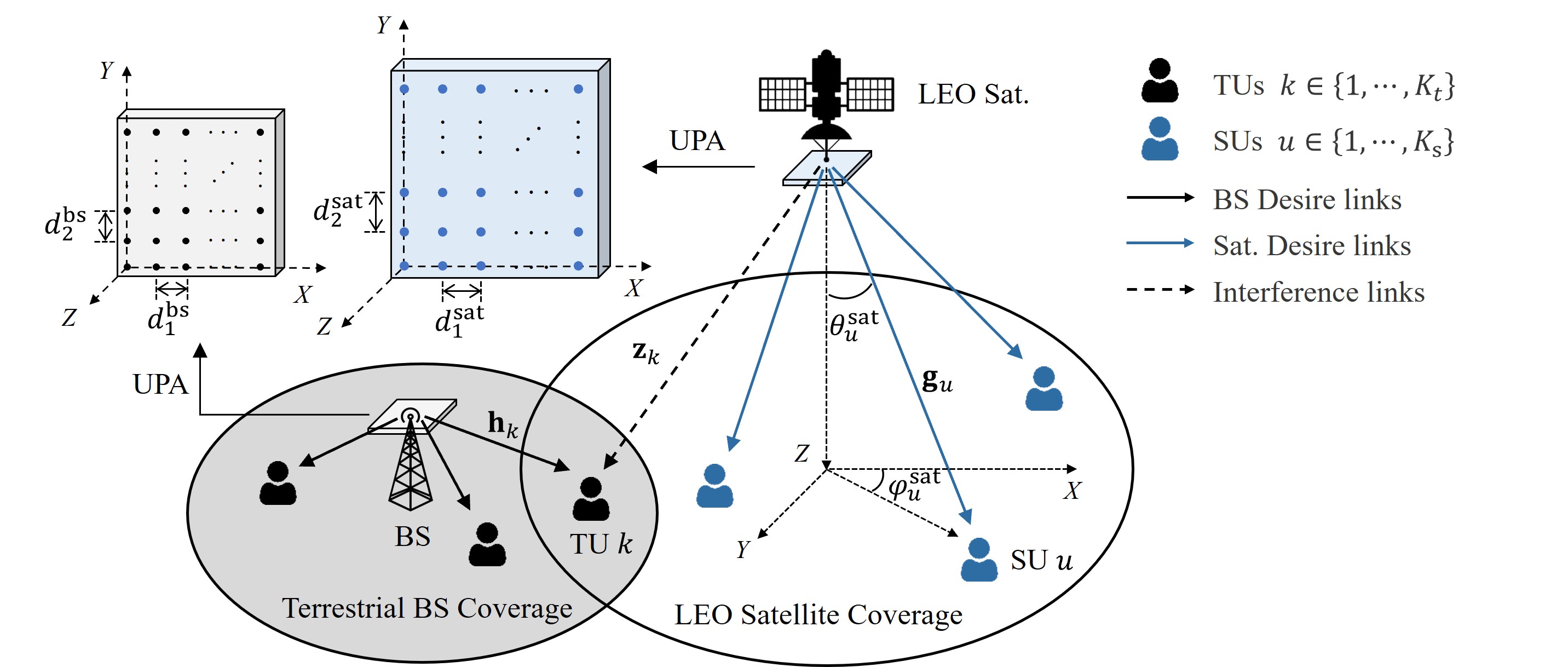}
    \caption{The system model of the STIN and the geometrical model of UPA.}
    \label{system model}
\end{figure}
 
We consider a STIN that consists of a LEO satellite and a terrestrial BS. We focus on the downlink system with full frequency reuse (FFR), wherein both the LEO satellite and the terrestrial BS use the same frequency band. 
As illustrated in Fig.\,\ref{system model}, there exist $K_s$ SUs and $K_t$ TUs, and the total number of users is $K = K_s + K_t$.
We denote $\CMcal{K}_s = \left\{1,\cdots,K_s\right\}$ as a set of the SUs with $\left|\CMcal{K}_s\right| = K_s$ and $\CMcal{K}_t=\left\{1,\cdots,K_t\right\}$ as a set of the TUs with $\left|\CMcal{K}_t\right| = K_t$.
In our setup, the terrestrial BS does not incur any interference to the SUs since all the SUs are located outside of the coverage region of the terrestrial BS\footnote{If the SUs enter the terrestrial coverage region, they change their association to the terrestrial BS and become TUs \cite{yin:twc:23}.}. On the contrary, the LEO satellite can incur interference to some TUs when they are located within the coverage region of the LEO satellite \cite{yin:twc:23} as shown in Fig.\,\ref{system model}.
We define a subset of the TUs that experiences the interference from the LEO satellite as $\CMcal{K}_t^{{\text{int}}} \subseteq \CMcal{K}_t$, where $|\CMcal{K}_t^{{\text{int}}}| = K_t^{{\text{int}}}$.
The total number of users in the LEO satellite coverage is $K^{\text{sat}} = K_t^{{\text{int}}} + K_s$.
If $K_t = K_s = 1$ and $\CMcal{K}_t^{\text{int}} = \CMcal{K}_t$, the considered setup corresponds to the Z channel \cite{chong:tit:07}. 
Our interference environment is an extension of the Z channel for $K_t \ge 1 $, $K_s \ge 1$, and $\CMcal{K}_t^{{\text{int}}} \subseteq \CMcal{K}_t$.

The LEO satellite is equipped with the uniform planar arrays (UPAs) with $M_1$ and $M_2$ array elements in the $x$-axis and $y$-axis, respectively. The total number of antennas at the LEO satellite is $M \triangleq M_1 M_2$. 
The terrestrial BS is also equipped with $N \triangleq N_1 N_2$ number of UPAs, where it consists of $N_1$ and $N_2$ array elements in the $x$-axis and $y$-axis, respectively.
We also assume that all users are equipped with a single antenna.

\subsection{Channel Model}

\subsubsection{Satellite Channel}
For modeling the satellite channel, we use a widely-adopted multi-path channel model. Using ray-tracing based modeling, the complex baseband channel impulse response of the downlink satellite channel vector ${\bf{g}}_{u}(t, f) \in \mathbb{C}^{M}$ for SU $u \in \CMcal{K}_s$ at instant $t$ and frequency $f$ is represented by 
\begin{align}
    {\bf{g}}_{u}(t, f) &= \frac{1}{\sqrt{L_s}} \sum_{\ell = 0}^{L_s -1} g_{u,\ell} e^{j 2\pi (\nu_{u, \ell} t - f \tau_{u, \ell})}\cdot {\bf{a}} ( \theta_{u, \ell}^{\text{sat}}, \varphi_{u, \ell}^{\text{sat}}, \CMcal{D}^{\text{sat}} ),
\end{align}
where $L_s$ is the number of propagation paths, $g_{u, \ell}$ is the complex channel gain, ${\nu}_{u, \ell}$ is the Doppler shift, $\tau_{u,\ell}$ is the propagation delay, and ${\bf{a}}(\cdot, \cdot, \cdot)$ is the array response vector corresponding to the $\ell$-th path of SU $u$'s channel. Note that $\theta_{u, \ell}^{\text{sat}}$ and $\varphi_{u, \ell}^{\text{sat}}$ represent the vertical and horizontal angle-of-departure (AoD) respectively, and $\CMcal{D}^{\text{sat}} = \{d_1^{\text{sat}}, d_2^{\text{sat}}\}$ is a set of the satellite's UPA inter-element spacing in the $x$-axis and $y$-axis. Without loss of generality, we assume that $\ell = 0$ indicates the first arriving path and $\ell = L_s - 1$ indicates the last arriving path at the satellite, resulting in $\tau_{u, 0} \le \tau_{u, 1} \le \cdots \le \tau_{u, L_s - 1}$.

{\bf{Doppler shift}}: We first explain the Doppler shift $\nu_{u, \ell}$. Typically, the Doppler shift for the LEO satellite channel is much larger compared to the terrestrial channel because of the high mobility of the LEO satellite. 
To deal with this, we can exploit a favorable characteristic of the LEO satellite channel. 
That is to say, each path has almost identical Doppler, i.e., $\nu_{u, \ell} = \nu_{u}$. This is because the traveling distances of each path can be assumed to be approximately the same due to the high altitude of the LEO satellite\footnote{
In practice, the Doppler shift can be varied over time due to moving direction of the satellite keeps changing. Taking into account this variation of Doppler shift is, however, beyond the scope of our paper. Thus we assume perfect knowledge on the Doppler shift.}.
This allows us to compensate the Doppler shift at the LEO satellite.
Note that the same Doppler shift compensation was also presented in \cite{You:jsac:20, Li:tcom:22, li:twc:23, you:twc:22}.

{\bf{Delay}}: The propagation delay is also an important issue in LEO satellite communications because of a long propagation distance. 
Similar to the Doppler shift compensation, we exploit the line-of-sight (LoS)-like characteristic of LEO satellite channels \cite{You:jsac:20, Li:tcom:22}.
Specifically, denoting the delay spread as $\tau^{\text{sp}}=\tau_{u, L_s-1} - \tau_{u, 0}$, $\tau^{\text{sp}}$ is rather smaller than that of terrestrial communications since the traveling distances between each propagation path are very similar.
We note that this is also confirmed with the measurement results \cite{vojcic:jsac:94, 3gpp:38:811}.
Therefore, if the receiver can achieve symbol synchronization for the minimum delay of $\tau_{u, 0}$, the delay spread can be readily resolved by using the typical orthogonal frequency division multiplex (OFDM) technique. 
This delay compensation was also employed in \cite{You:jsac:20, Li:tcom:22, li:twc:23, you:twc:22}.

{\bf{Effective channel}}: 
Incorporating the LoS-like characteristic of LEO satellite channel, we have ${\bf{a}}(\theta_{u, \ell}^{\text{sat}}, \varphi_{u, \ell}^{\text{sat}}, \CMcal{D}^{\text{sat}}) \simeq {\bf{a}}(\theta_{u}^{\text{sat}}, \varphi_{u}^{\text{sat}}, \CMcal{D}^{\text{sat}})$ for all $0 \le \ell \le L_{s}-1$. As a result, the effective LEO satellite channel that SU $u$ experiences in one OFDM symbol block is given by
\begin{align}
    {\bf{g}}_{u} = \tilde g_{u}\left( t, f \right) \cdot {\bf{a}} \left(\theta_{u}^{\text{sat}}, \varphi_{u}^{\text{sat}}, \CMcal{D}^{\text{sat}}\right), \label{sat_ch}
\end{align}
where $\tilde g_{u}\left(t, f\right) = \frac{1}{\sqrt{L_s}} \sum_{\ell = 0}^{L_s -1} g_{u,\ell} e^{j 2\pi (\nu_{u}t - f\tau_{u,0})}$ is the effective channel gain after Doppler and delay compensation, and ${\bf{a}} (\theta_{u}^{\text{sat}}, \varphi_{u}^{\text{sat}}, \CMcal{D}^{\text{sat}} )$ is the array response.
Since the LEO satellite channel is nearly LoS, not much spatial randomness exists in $\tilde g_{u}\left(t,f\right)$. To reflect this, we model $\tilde g_{u}\left(t,f\right)$ by Rician fading, i.e., $\tilde g_{u}\left(t,f\right) \sim \mathcal{CN}\left(\sqrt{\frac{\kappa_{s} \alpha_u}{1+\kappa_s}}, {\frac{\alpha_u}{1+\kappa_s}}\right)$, where $\alpha_u = \frac{G_{\textnormal{sat}} G_{u}}{k_{B} T_n B_w} \left(\frac{c}{4 \pi f_c d_0^{\text{sat}}}\right)^2$ is the channel power corresponding to SU $u$ with the free-space path loss. Here, $c$ is the speed of light, $f_c$ is the carrier frequency, $d_0^{\text{sat}}$ denotes the altitude of the LEO satellite, $k_{B}$ is the Boltzmann constant, $T_n$ is the noise temperature, $B_w$ is the system bandwidth, $G_{\textnormal{sat}}$ and $G_{u}$ respectively represent the antenna gains of the transmitter and the receiver, and $\kappa_{s}$ determines the ratio between the deterministic and random components. As $\kappa_{s} \rightarrow \infty$, $\tilde g_{u}\left(t,f\right)$ loses the randomness and becomes a deterministic constant, while $\kappa_{s} \rightarrow 0$, $\tilde g_{u}\left(t,f\right)$ is distributed as Rayleigh fading. We note that all of our channel modeling assumptions correspond to the prior work that studied the LEO satellite communications 
\cite{papa:personal:01, you:twc:22, You:jsac:20, Li:tcom:22}. The channel matrix between the LEO satellite and SUs is denoted by $\mathbf{G} = \left[\mathbf{g}_{1},\cdots,\mathbf{g}_{K_s}\right] \in \mathbb{C}^{M \times K_s}$.

Now we elaborate on the array response vector. In \eqref{sat_ch}, the array response vector ${\bf{a}} ( \theta_{u}^{\text{sat}}, \varphi_{u}^{\text{sat}}, \CMcal{D}^{\text{sat}} ) \in \mathbb{C}^{M}$ is given by
\begin{align}
    {\bf{a}} \left( \theta_{u}^{\text{sat}}, \varphi_{u}^{\text{sat}}, \CMcal{D}^{\text{sat}} \right) = \mathbf{a}_{h} \left(\theta_{u}^{\text{sat}},\varphi_{u}^{\text{sat}}, d_1^{\text{sat}} \right) \otimes \mathbf{a}_{v} \left(\theta_{u}^{\text{sat}}, d_2^{\text{sat}} \right),
\end{align}
where $\otimes$ denotes the Kronecker product, and the horizontal steering vector and the vertical steering vector are
$\mathbf{a}_{h}(\theta_{u}^{\text{sat}},\varphi_{u}^{\text{sat}}, d_1^{\text{sat}}) = \big[ e^{-j\frac{2\pi\left(M_1-1\right)}{2\lambda}d_1^{\text{sat}} \sin{\theta_{u}^{\text{sat}}}\cos{\varphi_{u}^{\text{sat}}}}, \cdots,  e^{+j\frac{2\pi\left(M_1-1\right)}{2\lambda}d_1^{\text{sat}} \sin{\theta_{u}^{\text{sat}}}\cos{\varphi_{u}^{\text{sat}}}} \big]^{\sf T} \in \mathbb{C}^{M_1}$ and $\mathbf{a}_{v}(\theta_{u}^{\text{sat}}, d_2^{\text{sat}}) = \big[e^{-j\frac{2\pi\left(M_2-1\right)}{2\lambda}d_2^{\text{sat}} \cos{\theta_{u}^{\text{sat}}}}, \cdots, e^{+j\frac{2\pi\left(M_2-1\right)}{2\lambda}d_2^{\text{sat}} \cos{\theta_{u}^{\text{sat}}}}\big]^{\sf T} \in \mathbb{C}^{M_2}$, respectively.
Here, $\lambda$ is the carrier wavelength.

Similarly, we also define the interfering channel from the LEO satellite to the TUs in $\CMcal{K}_t^{\text {int}}$. 
Applying the equivalent treatments to the Doppler shift and the delay, the effective interference channel experienced by TU $k \in \CMcal{K}_t^{{\text{int}}}$, denoted by $\mathbf{z}_{k}(t, f) \in \mathbb{C}^{M}$, is presented by
\begin{align}
    {\bf{z}}_{k}(t, f) &= \frac{1}{\sqrt{L_s}}\sum_{\ell = 0}^{L_s - 1} z_{k, \ell} e^{j 2\pi \left(\nu_{k, \ell}t - f\tau_{k, \ell}\right)} \cdot {\bf{a}}\left(\theta_{k, \ell}^{\text{sat}}, \varphi_{k, \ell}^{\text{sat}}, \CMcal{D}^{\text{sat}}\right) \nonumber\\
    &= \tilde z_{k}\left(t,f\right) \cdot {\bf{a}}\left(\theta_{k}^{\text{sat}}, \varphi_{k}^{\text{sat}}, \CMcal{D}^{\text{sat}}\right) = \mathbf{z}_{k}, \label{z_ch}
\end{align}
where $\tilde z_{k}\left(t,f\right)$ is the effective channel gain, which follows $\tilde z_{k}\left(t,f\right) \sim \mathcal{CN}\left(\sqrt{\frac{\kappa_{s} \alpha_k}{1+\kappa_s}}, {\frac{\alpha_k}{1+\kappa_s}}\right)$. 
The interfering channel matrix between the LEO satellite and TUs in $\CMcal{K}_t^{\text {int}}$ is defined as $\mathbf{Z} = \big[ \mathbf{z}_{1}, \cdots, \mathbf{z}_{K_t^{{\text{int}}}} \big] \in \mathbb{C}^{M \times K_t^{{\text{int}}}}$.

\subsubsection{Terrestrial Channel}

In the terrestrial channel, we assume there is no LoS component by considering dense urban scenarios where large-scale blockages such as buildings are densely placed. 
Assuming that there are $L_t$ scatters contributing the Non-LoS components, the channel vector ${\bf{h}}_{k} \in \mathbb{C}^{N}$ between the terrestrial BS and TU $k$ is given by 
\begin{align}
    \mathbf{h}_{k} = {\frac{1}{\sqrt L_t}}\sum_{\ell=0}^{L_t-1} h_{k,\ell} \cdot \mathbf{a}_{h}\left(\theta_{k,\ell}^{\text{bs}},\varphi_{k,\ell}^{\text{bs}}, d_1^{\text{bs}}\right) \otimes \mathbf{a}_{v}\left(\theta_{k,\ell}^{\text{bs}}, d_2^{\text{bs}}\right),\label{BS_ch}
\end{align}
where $h_{k,\ell}$ is the complex channel gain of the $\ell$-th path, which follows $\mathcal{CN}\left( 0,\beta_k \right)$ with $\beta_{k} = \frac{G_{\textnormal{bs}} G_{k} }{k_{B} T_n B_w} \left(\frac{c}{4 \pi f_c }\right)^2 \Big(\frac{1}{d_k^{\text{bs}}}\Big)^{\rho}$. Here, $d_k^{\text{bs}}$ is the distance between the terrestrial BS and TU $k$, and $\rho$ is the path loss exponent.
We denote that $\theta_{k,\ell}^{\text{bs}}$ and $\varphi_{k,\ell}^{\text{bs}}$ as the vertical AoD and the horizontal AoD for the $\ell$-th path and $\CMcal{D}^{\text{bs}} = \{d_1^{\text{bs}}, d_2^{\text{bs}}\}$ as a set of the terrestrial BS's UPA inter-element spacing in the $x$-axis and $y$-axis, respectively. In \eqref{BS_ch}, the horizontal steering vector and the vertical steering vector are denoted by $\mathbf{a}_{h}(\theta_{k,\ell}^{\text{bs}},\varphi_{k,\ell}^{\text{bs}},d_1^{\text{bs}}) = \big[e^{-j\frac{2\pi\left(N_1-1\right)}{2\lambda}d_1^{\text{bs}} \sin{\theta_{k,\ell}^{\text{bs}}}\cos{\varphi_{k,\ell}^{\text{bs}}}}, \cdots,  e^{+j\frac{2\pi\left(N_1-1\right)}{2\lambda}d_1^{\text{bs}} \sin{\theta_{k,\ell}^{\text{bs}}}\cos{\varphi_{k,\ell}^{\text{bs}}}}\big]^{\sf T} \in \mathbb{C}^{N_1}$ and $\mathbf{a}_{v} (\theta_{k,\ell}^{\text{bs}},d_2^{\text{bs}}) = \big[e^{-j\frac{2\pi\left(N_2-1\right)}{2\lambda}d_2^{\text{bs}} \cos{\theta_{k,\ell}^{\text{bs}}}}, \cdots, e^{+j\frac{2\pi\left(N_2-1\right)}{2\lambda}d_2^{\text{bs}} \cos{\theta_{k,\ell}^{\text{bs}}}}\big]^{\sf T} \in \mathbb{C}^{N_2}$, respectively.
The terrestrial channel matrix is $\mathbf{H} = \left[\mathbf{h}_{1},\cdots,\mathbf{h}_{K_t} \right] \in \mathbb{C}^{N \times K_t}$.
As mentioned above, the channels from the terrestrial BS to the SUs are negligible because the SUs are located outside of the terrestrial BS coverage region.

\subsubsection{CSIT Estimation}
Now we explain the CSIT estimation. We assume that the perfect CSI is given at the users (including the SUs and the TUs) (i.e., perfect CSI at the receiver (CSIR))\footnote{
In practice, channel estimation errors also can occur at the receivers. For the sake of conciseness, this paper does not consider the case of imperfect CSIR. Incorporating imperfect CSIR is promising as future research.}.
Assuming time division duplex (TDD) that allows the channel reciprocity, each user sends a pilot sequence to estimate the uplink channel. Thanks to the reciprocity, the estimated uplink CSI is reused in design of downlink precoders. In this case, the CSIT estimation error is modeled as a function of the uplink pilot power and pilot sequence length \cite{Zheng:jsac:22}.
To be specific, we assume that the CSIT is estimated via linear MMSE at both the LEO satellite and the terrestrial BS.
We denote the estimated channel vectors $\{\hat{\mathbf{g}}_{u}, \hat{\mathbf{z}}_{k}\}\in\mathbb{C}^{M}$ and $\hat{\mathbf{h}}_{k}\in\mathbb{C}^{N}$ as 
\begin{align}
    \hat{\mathbf{g}}_{u} = \mathbf{g}_{u} - \mathbf{q}_{u}^{\text{sat}},\; \hat{\mathbf{z}}_{k} =  \mathbf{z}_{k} - \mathbf{e}_{k}^{\text{sat}},\; \hat{\mathbf{h}}_{k} = \mathbf{h}_{k} - \mathbf{e}_{k}^{\text{bs}},\label{h_hat}
\end{align}
where $\{\mathbf{q}_{u}^{\text{sat}},\mathbf{e}_{k}^{\text{sat}}\}\in\mathbb{C}^{M}$ and $\mathbf{e}_{k}^{\text{bs}}\in\mathbb{C}^{N}$ are the CSIT estimation error vectors.
The error covariance matrices for each channel are given by a function of the channel covariance matrices, the pilot length $\tau$, the pilot transmission power $p^{\textnormal{pi}}$, and the estimation noise power $\sigma^2$:
\begin{align}
    \mathbf{\Psi}_{u}^{\textnormal{sat}} &= \mathbb{E}\left[\mathbf{q}_{u}^{\textnormal{sat}}(\mathbf{q}_{u}^{\textnormal{sat}})^{\sf H}\right] = \mathbf{Q}_{u} - \mathbf{Q}_{u} \left(\mathbf{Q}_{u} + \frac{\sigma^2}{\tau p^{\textnormal{pi}}}\mathbf{I}_{M}\right)^{-1}\mathbf{Q}_{u},\label{Q_SU}\\
    \mathbf{\Phi}_{k}^{\textnormal{sat}} &= \mathbb{E}\left[\mathbf{e}_{k}^{\textnormal{sat}}(\mathbf{e}_{k}^{\textnormal{sat}})^{\sf H}\right] = \mathbf{R}_{k}^{\textnormal{sat}} - \mathbf{R}_{k}^{\textnormal{sat}} \left(\mathbf{R}_{k}^{\textnormal{sat}} + \frac{\sigma^2}{\tau p^{\textnormal{pi}}}\mathbf{I}_{M}\right)^{-1}\mathbf{R}_{k}^{\textnormal{sat}},\label{R_TU}\\
    \mathbf{\Phi}_{k}^{\textnormal{bs}} &= \mathbb{E}\left[\mathbf{e}_{k}^{\textnormal{bs}}(\mathbf{e}_{k}^{\textnormal{bs}})^{\sf H}\right] = \mathbf{R}_{k}^{\textnormal{bs}} - \mathbf{R}_{k}^{\textnormal{bs}} \left(\mathbf{R}_{k}^{\textnormal{bs}} + \frac{\sigma^2}{\tau p^{\textnormal{pi}}}\mathbf{I}_{N}\right)^{-1}\mathbf{R}_{k}^{\textnormal{bs}},\label{R_SU}
\end{align}
where $\mathbf{Q}_{u} = \mathbb{E}[\mathbf{g}_{u}\mathbf{g}_{u}^{\sf H}] \in\mathbb{C}^{M \times M}, \mathbf{R}_{k}^{\textnormal{sat}} = \mathbb{E}[\mathbf{z}_{k}\mathbf{z}_{k}^{\sf H}] \in\mathbb{C}^{M \times M}$, and $\mathbf{R}_{k}^{\textnormal{bs}} = \mathbb{E}[\mathbf{h}_{k}\mathbf{h}_{k}^{\sf H}] \in\mathbb{C}^{N \times N}$ are the channel spatial covariance matrices, and $\mathbf{I}_M \in\mathbb{C}^{M \times M}, \mathbf{I}_N \in\mathbb{C}^{N \times N}$ are the identity matrices.
We assume that the channel spatial covariance matrices are known at the LEO satellite and the terrestrial BS by exploiting the long-term channel information such as AoDs\footnote{Due to high mobility of LEO satellites, however, even this long-term channel information can be frequently varying compared to that of terrestrial channels. Incorporating this into precoder design is interesting future work.}  \cite{You:jsac:20, Li:tcom:22}. 

Our model generalizes the previous satellite CSIT acquisition model. 
For example, in \cite{You:jsac:20}, two cases of CSIT acquisition were considered, instantaneous CSI (iCSI) and statistical CSI (sCSI). 
In iCSI, it is assumed that the LEO satellite has perfect knowledge of instantaneous CSIT, i.e., $\mathbf{q}_{u}^{\text{sat}} = \mathbf{e}_{k}^{\text{sat}} = 0$ in \eqref{h_hat}. We note that this is achieved by $\tau p^{\textnormal{pi}} \rightarrow \infty$, so that the channel estimation error becomes negligible. 
On the contrary to that, sCSI assumes that only the long-term channel covariance is available without any instantaneous channel knowledge. Since our model assumes that the channel covariance matrices are known, the sCSI case corresponds to 
$ \tau p^{\textnormal{pi}}= 0$, i.e., no instantaneous channel estimation is used. 
In this sense, our model encompasses iCSI and sCSI as two extreme cases.
In practice, it is more plausible that CSIT is partially known with some error depending on the channel estimation methods as in our model.

Lastly, we emphasize that the proposed distributed precoding method is applicable not only in the considered CSIT estimation model, but also in an arbitrary CSIT estimation model. To be specific, the proposed method is developed based on a lower bound of the spectral efficiency, characterized by using the CSIT error covariance. For this reason, given that the CSIT error covariance is attainable, it is feasible to use the proposed precoding method in any CSIT estimation models. 
We will explain it in more details in Section \uppercase\expandafter{\romannumeral4}.

\subsection{Signal Model}
{\bf{Satellite transmit signal}}: To effectively handle the interference of the LEO satellite, we employ the RS strategy. Following the principle of 1-layer RS \cite{park:twc:23, Li:jsac:20}, the message $M_u$, intended to SU $u$, is split into a common part and a private part, i.e., $M_{u} \rightarrow \{M_{c,u}, M_{p,u}\}$. The common parts of all the messages $\{ M_{c,1},\cdots,M_{c,K_s} \}$ are jointly combined and encoded into a common stream $s_{c}$. The private parts $M_{p,u}$ are independently encoded into the private stream $s_{p,u}$. For the common stream $s_{c}$, the corresponding codebook is shared with both of the SUs in $\CMcal{K}_s$ and the TUs in $\CMcal{K}_{t}^{{\text{int}}}$, who experience the interference from the LEO satellite. The codebook for the private stream $s_{p,u}$ is only given to SU $u$. Accordingly, $s_{c}$ is decodable to the users in $\CMcal{K}_s$ and $\CMcal{K}_t^{{\text{int}}}$, while $s_{p,u}$ is decodable only to SU $u$. Note that $s_{c}$ and $s_{p,u}$ are drawn from independent Gaussian codebooks, i.e., $s_{c}, s_{p,u} \sim \mathcal{CN}(0,P_s)$ where $P_s$ is the total transmit power of the LEO satellite.

At the LEO satellite, the common stream $s_{c}$ and the private streams $s_{p,u}$ are linearly combined with the precoding vectors $\mathbf{f}_{c} \in\mathbb{C}^{M}$ and $\mathbf{f}_{p,u} \in\mathbb{C}^{M}$. Then the transmit signal vector of the LEO satellite is defined as $\mathbf{x}^{\textnormal{sat}} = \mathbf{f}_{c} s_{c} + \sum_{i=1}^{K_s} \mathbf{f}_{p,i} s_{p,i} \in\mathbb{C}^{M}$.
We let $\mathbf{F} = \left[\mathbf{f}_c, \mathbf{f}_{p,1}, \cdots, \mathbf{f}_{p,K_s}\right] \in\mathbb{C}^{M \times \left(K_s+1\right)}$ as the precoding matrix of the LEO satellite. For the transmit power constraint, we assume $\textnormal{tr}(\mathbf{F}\mathbf{F}^{\sf H}) \leq 1$, by which the total transmit power is constrained by $P_s$.

{\bf{Terrestrial BS transmit signal}}: In the terrestrial BS, the messages $W_{1},\cdots,W_{K_t}$ are directly encoded into the streams $m_{1},\cdots,m_{K_t}$ without RS. 
Each stream $m_{k}$ is drawn from an independent Gaussian codebook, i.e., $m_{k} \sim \mathcal{CN}\left(0,P_t\right)$ where $P_t$ is the total transmit power of the terrestrial BS. 
Then, the streams are linearly combined with the precoding vectors ${\bf{v}}_{k} \in \mathbb{C}^{N}$. 
Accordingly, the transmit signal of the terrestrial BS is given by $\mathbf{x}^{\textnormal{bs}} = \sum_{j=1}^{K_t} \mathbf{v}_{j} m_{j} \in \mathbb{C}^{N}$.
The precoding matrix of the terrestrial BS is $\mathbf{V} = \left[\mathbf{v}_{1}, \cdots, \mathbf{v}_{K_t}\right] \in\mathbb{C}^{N \times K_t}$ and the power constraint of terrestrial BS is $\textnormal{tr}(\mathbf{V}\mathbf{V}^{\sf H}) \leq 1$.

{\bf{Received signal}\footnote{With compensation for Doppler shift and delay at each user, we assume perfect synchronization in both time and frequency between the terrestrial BS and TUs $\forall k \in \CMcal{K}_t$, as well as between the LEO satellite and both SUs $\forall u \in \CMcal{K}_s$ and TUs $\forall k \in \CMcal{K}_t^{{\text{int}}}$ \cite{You:jsac:20,Li:tcom:22}.} }: 
The received signal at TU $k \in \CMcal{K}_t^{\text{int}}$ is 
\begin{align} 
    y_{k} = \mathbf{h}_{k}^{\sf H}\sum_{j=1}^{K_t} \mathbf{v}_{j} m_{j} + \mathbf{z}_{k}^{\sf H} \left( \mathbf{f}_{c} s_{c} + \sum_{i=1}^{K_s} \mathbf{f}_{p,i} s_{p,i} \right) + n_{k},\label{rx_bs}
\end{align}
where $n_{k}$ is the additive white Gaussian noise (AWGN) with variance $\sigma^2$ and ${\bf{z}}_k$ is the interference channel from the LEO satellite to TU $k$. For the received signal at TU $k \notin \CMcal{K}_t^{\text{int}}$, we have ${\bf{z}}_k = {\bf{0}}$ in \eqref{rx_bs}. 
The received signal at SU $u$ is 
\begin{align}
    y_{u} = \mathbf{g}_{u}^{\sf H} \left( \mathbf{f}_{c} s_{c} + \sum_{i=1}^{K_s} \mathbf{f}_{p,i} s_{p,i} \right) + n_{u},\label{rx_sat}
\end{align}
where $n_{u} \sim \mathcal{CN}(0,\sigma^2)$ is AWGN. 

\subsection{Performance Metrics and Problem Formulation}
\subsubsection{Spectral Efficiency Characterization}
Before formulating our main problem, we explain the decoding process in the considered RS strategy. The SUs in $\CMcal{K}_s$ and the TUs in $\CMcal{K}_t^{{\text{int}}}$ first decode the common stream $s_{c}$ by treating all the other private streams as noise. 
After successfully decoding the common stream, each user removes the common stream from the received signal by using SIC. Then, SU $u$ and TU $k$ decode $s_{p,u}$ and $m_{k}$ respectively, by treating any residual interference as noise.

The SINR of the common stream $s_{c}$ for TU $k \in \CMcal{K}_t^{{\text{int}}}$ and SU $u$ are respectively given as
\begin{align}
    &\textnormal{SINR}_{c,k \in \CMcal{K}_t^{{\text{int}}}}^{\textnormal{bs}} = \frac{ \left|\mathbf{z}_{k}^{\sf H} \mathbf{f}_{c}\right|^2}{\frac{P_t}{P_s}\sum_{j=1}^{K_t} \left|\mathbf{h}_{k}^{\sf H} \mathbf{v}_{j}\right|^2 + \sum_{i=1}^{K_s} \left|\mathbf{z}_{k}^{\sf H} \mathbf{f}_{p,i}\right|^2 + \frac{\sigma^2}{P_s}},\label{SINR_com_TU}\\
    &\textnormal{SINR}_{c,u}^{\textnormal{sat}} = \frac{\left|\mathbf{g}_{u}^{\sf H}\mathbf{f}_{c}\right|^2}{\sum _{i=1}^{K_s} \left|\mathbf{g}_{u}^{\sf H} \mathbf{f}_{p,i}\right|^2 + \frac{\sigma^2}{P_s}}.\label{SINR_com_SU}
\end{align}
Based on these, the spectral efficiencies are represented as $R_{c,k}^{\textnormal{bs}}\left(\mathbf{F},\mathbf{V}\right) = \log_2 \big(1 + \textnormal{SINR}_{c,k}^{\textnormal{bs}}\big)$ and $R_{c,u}^{\textnormal{sat}}\left(\mathbf{F}\right) = \log_2 \big(1 + \textnormal{SINR}_{c,u}^{\textnormal{sat}}\big)$, respectively.

To guarantee that the users in $\CMcal{K}_t^{{\text{int}}}$ and $\CMcal{K}_s$ can decode the common stream $s_{c}$, the code rate of the common stream should be determined as the minimum value of the spectral efficiencies among the users in $\CMcal{K}_t^{{\text{int}}}$ and $\CMcal{K}_s$; thus, $R_{c}\left(\mathbf{F},\mathbf{V}\right) = \min_{k \in \CMcal{K}_t^{{\text{int}}}, u \in \CMcal{K}_s} \big\{R_{c,k}^{\textnormal{bs}}\left(\mathbf{F},\mathbf{V}\right), R_{c,u}^{\textnormal{sat}}\left(\mathbf{F}\right)\big\}$.
We clarify that, under the assumption that 
the code rate of the common stream $s_c$ is set to be less than or equal to $R_{c}\left(\mathbf{F},\mathbf{V}\right)$, the decoding error for $s_{c}$ exponentially decreases as the code length increases. 
The SINR of the private stream $m_{k}$ for the TUs in $\CMcal{K}_t$ is given by 
\begin{align}
    &\textnormal{SINR}_{p,k \in \CMcal{K}_t^{{\text{int}}}}^{\textnormal{bs}} = \frac{\left|\mathbf{h}_{k}^{\sf H}\mathbf{v}_{k}\right|^2}{\sum_{j=1, j \neq k}^{K_t} \left|\mathbf{h}_{k}^{\sf H} \mathbf{v}_{j}\right|^2 + \frac{P_s}{P_t}\sum_{i=1}^{K_s} \left|\mathbf{z}_{k}^{\sf H} \mathbf{f}_{p,i}\right|^2 + \frac{\sigma^2}{P_t}},\label{SINR_pri_TU}\\
    &{\textnormal{SINR}_{p,k \notin \CMcal{K}_t^{{\text{int}}}}^{\textnormal{bs}} = \frac{\left|\mathbf{h}_{k}^{\sf H}\mathbf{v}_{k}\right|^2}{\sum_{j=1, j \neq k}^{K_t} \left|\mathbf{h}_{k}^{\sf H} \mathbf{v}_{j}\right|^2 + \frac{\sigma^2}{P_t}}.}\label{SINR_pri_nTU}
\end{align}
For the sake of brevity, we henceforth assume ${\bf{z}}_k = 0$ for TU $k \notin \CMcal{K}_t^{{\text{int}}}$ and express the performance of all TUs in a unified manner. 
The SINR of the private stream $s_{p,u}$ for SU $u$ is
\begin{align}
    \textnormal{SINR}_{p,u}^{\textnormal{sat}} = \frac{\left|\mathbf{g}_{u}^{\sf H}\mathbf{f}_{p,u}\right|^2}{\sum_{i=1, i \neq u}^{K_s} \left|\mathbf{g}_{u}^{\sf H} \mathbf{f}_{p,i}\right|^2 + \frac{\sigma^2}{P_s}}.\label{SINR_pri_SU}
\end{align}
The spectral efficiencies of the private streams $m_{k}$ and $s_{p,u}$ are respectively $R_{p,k}^{\textnormal{bs}}\left(\mathbf{F},\mathbf{V}\right) = \log_2 \big(1 + \textnormal{SINR}_{p,k}^{\textnormal{bs}}\big)$ and $R_{p,u}^{\textnormal{sat}}\left(\mathbf{F}\right) = \log_2 \big(1 + \textnormal{SINR}_{p,u}^{\textnormal{sat}}\big)$.

Unfortunately, however, the spectral efficiencies $R_{c}({\bf{F}}, {\bf{V}}), R_{p,k}^{\textnormal{bs}}\left(\mathbf{F},\mathbf{V}\right)$, and $R_{p,u}^{\textnormal{sat}}\left(\mathbf{F}\right)$ cannot be computed at the transmitters (the terrestrial BS and the LEO satellite) under an imperfect CSIT setup. To address this, we derive a lower bound on the spectral efficiency. To this end, we rewrite the received signals \eqref{rx_bs} and \eqref{rx_sat} incorporating the CSIT estimation error as 
{\color{black}{
\begin{align}
    y_{k} &= \left( \hat{\mathbf{h}}_{k} + \mathbf{e}_{k}^{\textnormal{bs}} \right)^{\sf H} \sum_{j=1}^{K_t} \mathbf{v}_{j} m_{j} + \left( \hat{\mathbf{z}}_{k} + \mathbf{e}_{k}^{\textnormal{sat}} \right)^{\sf H} \left( \mathbf{f}_{c} s_{c} + \sum_{i=1}^{K_s} \mathbf{f}_{p,i} s_{p,i} \right) \nonumber\\
    &+ n_{k},\label{rx_bs_low}\\
    y_{u} &= \left( \hat{\mathbf{g}}_{u} + \mathbf{q}_{u}^{\textnormal{sat}} \right)^{\sf H} \left( \mathbf{f}_{c} s_{c} + \sum_{i=1}^{K_s} \mathbf{f}_{p,i} s_{p,i} \right) + n_{u}.\label{rx_sat_low}
\end{align}
}}
Applying a generalized mutual information (GMI) technique \cite{Ding:tit:10}, 
we treat the CSIT estimation error as independent Gaussian noise with appropriate moment matching. 
This makes a lower bound on $R_{c,k}^{\textnormal{bs}}\left(\mathbf{F},\mathbf{V}\right)$ as follows:
{\color{black}{
\begin{align}
    &R_{c,k}^{\textnormal{bs}}\left(\mathbf{F},\mathbf{V}\right) 
    \overset{(a)}{\geq} \mathbb{E} \left[ \log_2 \left(1 + \frac{\left|\hat{\mathbf{z}}_{k}^{\sf H} \mathbf{f}_{c}\right|^2}{ I_{c,k}^{U} + I_{c,k}^{C} + \lvert(\mathbf{e}_{k}^{\textnormal{sat}})^{\sf H} \mathbf{f}_c \rvert^2 + \frac{\sigma^2}{P_s} }\right) \right] \nonumber\\
    &\overset{(b)}{\geq} \log_2\left(1 + \frac{\left|\hat{\mathbf{z}}_{k}^{\sf H} \mathbf{f}_{c}\right|^2}{ \tilde{I}_{c,k}^{U} + \tilde{I}_{c,k}^{C} + \mathbf{f}_c^{\sf H}\mathbb{E}\left[\mathbf{e}_{k}^{\textnormal{sat}}(\mathbf{e}_{k}^{\textnormal{sat}})^{\sf H}\right]\mathbf{f}_c + \frac{\sigma^2}{P_s} } \right) \label{low_R_ck}\\
    &\overset{(c)}{=} \log_2 \left(1 + \frac{\left|\hat{\mathbf{z}}_{k}^{\sf H} \mathbf{f}_{c} \right|^2}{\bar{I}_{c,k}^{U} + \bar{I}_{c,k}^{C} + \mathbf{f}_c^{\sf H}\mathbf{\Phi}_{k}^{\textnormal{sat}}\mathbf{f}_c + \frac{\sigma^2}{P_s}} \right) \nonumber\\
    &\,= \Bar{R}_{c,k}^{\textnormal{bs}}\left(\mathbf{F},\mathbf{V}\right),\label{R_com_TU_lower}
\end{align}
where $I_{c,k}^{U} = \frac{P_t}{P_s} \sum_{j=1}^{K_t} \lvert (\hat{\mathbf{h}}_{k} + \mathbf{e}_{k}^{\textnormal{bs}})^{\sf H} \mathbf{v}_{j} \rvert^2$ and $I_{c,k}^{C} = \sum_{i=1}^{K_s} \lvert (\hat{\mathbf{z}}_{k} + \mathbf{e}_{k}^{\textnormal{sat}})^{\sf H} \mathbf{f}_{p,i} \rvert^2$.
In $(a)$, the expectation is taken over the randomness associated with the CSIT estimation error $\{\mathbf{e}_{k}^{\textnormal{bs}},\mathbf{e}_{k}^{\textnormal{sat}}\}$.
$(b)$ is obtained by applying Jensen's inequality where $\tilde{I}_{c,k}^{U} = \frac{P_t}{P_s} \sum_{j=1}^{K_t} ( \lvert \hat{\mathbf{h}}_{k}^{\sf H}\mathbf{v}_{j} \rvert^2 + \mathbf{v}_{j}^{\sf H}\mathbb{E} [\mathbf{e}_{k}^{\textnormal{bs}}(\mathbf{e}_{k}^{\textnormal{bs}})^{\sf H} ]\mathbf{v}_{j} )$ and $\tilde{I}_{c,k}^{C} = \sum_{i=1}^{K_s} ( \lvert \hat{\mathbf{z}}_{k}^{\sf H} \mathbf{f}_{p,i} \rvert^2 + \mathbf{f}_{p,i}^{\sf H} \mathbb{E} [\mathbf{e}_{k}^{\textnormal{sat}}(\mathbf{e}_{k}^{\textnormal{sat}})^{\sf H} ] \mathbf{f}_{p,i} )$.
In $(c)$, by applying \eqref{R_TU} and \eqref{R_SU} to \eqref{low_R_ck}, as evident and in \eqref{R_com_TU_lower}, a lower bound on $R_{c,k}^{\textnormal{bs}}\left(\mathbf{F},\mathbf{V}\right)$ is given as $\Bar{R}_{c,k}^{\textnormal{bs}}\left(\mathbf{F},\mathbf{V}\right)$ where $\bar{I}_{c,k}^{U}=\frac{P_t}{P_s}\sum_{j=1}^{K_t} ( \lvert \hat{\mathbf{h}}_{k}^{\sf H}\mathbf{v}_{j} \rvert^2 + \mathbf{v}_{j}^{\sf H}\mathbf{\Phi}_{k}^{\textnormal{bs}}\mathbf{v}_{j} )$ and $\bar{I}_{c,k}^{C}=\sum_{i=1}^{K_s} ( \lvert \hat{\mathbf{z}}_{k}^{\sf H} \mathbf{f}_{p,i} \rvert^2 + \mathbf{f}_{p,i}^{\sf H}\mathbf{\Phi}_{k}^{\textnormal{sat}}\mathbf{f}_{p,i} )$.}}
Since the error covariance matrices $\mathbf{\Phi}_{k}^{\textnormal{bs}}$ and $\mathbf{\Phi}_{k}^{\textnormal{sat}}$ are assumed to be known at the LEO satellite, we are able to calculate \eqref{R_com_TU_lower} as a closed-form. 
Likewise, a lower bound on the spectral efficiency of the common stream at SU $u$ is acquired as follows.
{\color{black}
{\begin{align} 
    R_{c,u}^{\textnormal{sat}}\left(\mathbf{F}\right) 
    &\geq \log_2\left(1 + \frac{\left|\hat{\mathbf{g}}_{u}^{\sf H} \mathbf{f}_{c}\right|^2}{\bar{I}_{c,u}^{U} + \mathbf{f}_c^{\sf H}\mathbf{\Psi}_{u}^{\textnormal{sat}}\mathbf{f}_c + \frac{\sigma^2}{P_s}} \right) \nonumber\\
    &= \Bar{R}_{c,u}^{\textnormal{sat}}\left(\mathbf{F}\right),\label{R_com_SU_lower}
\end{align}
where $\bar{I}_{c,u}^{U}=\sum_{i=1}^{K_s} ( \lvert\hat{\mathbf{g}}_{u}^{\sf H}\mathbf{f}_{p,i}\rvert^2 + \mathbf{f}_{p,i}^{\sf H}\mathbf{\Psi}_{u}^{\textnormal{sat}}\mathbf{f}_{p,i} )$.
Similar to this, we obtain a lower bound on the spectral efficiency of the private stream for TU $k$ and SU $u$ as 
\begin{align}
    R_{p,k}^{\textnormal{bs}}\left(\mathbf{F},\mathbf{V}\right) 
    &\ge \log_2\bigg(1 + \frac{\left|\hat{\mathbf{h}}_{k}^{\sf H}\mathbf{v}_{k}\right|^2}{\bar{I}_{p,k}^{U} + \bar{I}_{p,k}^{C} + \mathbf{v}_{k}^{\sf H}\mathbf{\Phi}_{k}^{\textnormal{bs}}\mathbf{v}_{k}  + \frac{\sigma^2}{P_t}}\bigg) \nonumber\\
    &= \Bar{R}_{p,k}^{\textnormal{bs}}\left(\mathbf{F},\mathbf{V}\right),\label{R_pri_TU_lower}\\
    R_{p,u}^{\textnormal{sat}}\left(\mathbf{F}\right)
    &\ge \log_2\left(1 + \frac{\left|\hat{\mathbf{g}}_{u}^{\sf H}\mathbf{f}_{p,u}\right|^2}{\bar{I}_{p,u}^{U} + \mathbf{f}_{p,u}^{\sf H}\mathbf{\Psi}_{u}^{\textnormal{sat}}\mathbf{f}_{p,u} + \frac{\sigma^2}{P_s} } \right) \nonumber\\
    &= \Bar{R}_{p,u}^{\textnormal{sat}}\left(\mathbf{F}\right),\label{R_pri_SU_lower}
\end{align}
where $\bar{I}_{p,k}^{U} = \sum_{j=1, j \neq k}^{K_t} ( \lvert \hat{\mathbf{h}}_{k}^{\sf H}\mathbf{v}_{j} \rvert^2 +  \mathbf{v}_{j}^{\sf H}\mathbf{\Phi}_{k}^{\textnormal{bs}}\mathbf{v}_{j} )$, $\bar{I}_{p,k}^{C} = \frac{P_s}{P_t} \sum_{i=1}^{K_s} ( \lvert \hat{\mathbf{z}}_{k}^{\sf H} \mathbf{f}_{p,i} \rvert^2 + \mathbf{f}_{p,i}^{\sf H}\mathbf{\Phi}_{k}^{\textnormal{sat}}\mathbf{f}_{p,i} )$ and $\bar{I}_{p,u}^{U}=\sum_{i=1, i \neq u}^{K_s} ( \lvert\hat{\mathbf{g}}_{u}^{\sf H}\mathbf{f}_{p,i} \rvert^2 +  \mathbf{f}_{p,i}^{\sf H}\mathbf{\Psi}_{u}^{\textnormal{sat}}\mathbf{f}_{p,i} )$.}}
With the obtained lower bounds on the spectral efficiencies, we formulate our main problem as follows.

\subsubsection{Problem Formulation}
We formulate the following precoder design problem that aims to maximize a lower bound on the sum spectral efficiency with the RS strategy: 
\begin{align}
    \mathscr{P}_1: \; &\underset{\left\{\mathbf{F}, \mathbf{V}\right\}}{\textnormal{maximize}}
    \;\; 
    {\color{black}{
    \begin{bmatrix}
        \min_{k \in \CMcal{K}_t^{{\text{int}}}, u \in \CMcal{K}_s} \left\{\Bar{R}_{c,k}^{\textnormal{bs}}\left(\mathbf{F},\mathbf{V}\right), \Bar{R}_{c,u}^{\textnormal{sat}}\left(\mathbf{F}\right)\right\}\\
        + \sum_{i=1}^{K_s} \Bar{R}_{p,i}^{\textnormal{sat}}\left(\mathbf{F}\right) + \sum_{j=1}^{K_t} \Bar{R}_{p,j}^{\textnormal{bs}}\left(\mathbf{F},\mathbf{V}\right)
    \end{bmatrix}
    }},\label{opt_prob}\\
    &\textnormal{subject to} \quad \textnormal{tr}\left(\mathbf{F}\mathbf{F}^{\sf H}\right) = \lVert\mathbf{f}_c\rVert^2 + \sum_{i=1}^{K_s}\lVert\mathbf{f}_{p,i}\rVert^2 \leq 1,\nonumber\\
    &\quad\quad\quad\quad\quad\textnormal{tr}\left(\mathbf{V}\mathbf{V}^{\sf H}\right) = \sum_{j=1}^{K_t}\lVert\mathbf{v}_{j}\rVert^2 \leq 1.
\end{align}
We summarize the challenges for solving $\mathscr{P}_1$ as follows.
\lowercase\expandafter{\romannumeral1}) The estimated CSIT should be shared between the LEO satellite the terrestrial BS, which is difficult in the considered STINs. 
\lowercase\expandafter{\romannumeral2}) $\mathscr{P}_1$ is non-convex and also non-smooth, wherein finding a global optimal solution within polynomial time is infeasible. 
In the next section, we resolve these difficulties.

{\color{black}{
\begin{remark} \normalfont
(Application of Jensen's inequality)
    If the desired signal power and the interference signal power are independent, applying Jensen's inequality yields a lower bound as in \eqref{low_R_ck}. This is because $f(I) = \log_2 (1+ c/I)$ is convex. 
    As the number of antennas $M$ and $N$ increase, the gap between the exact rate expression and the approximation obtained by Jensen's inequality becomes negligible even for a case where the desired signal and the interference signal are not necessarily independent as shown in \cite{zhang:jstsp:14}. 
    For a more comprehensive explanation, we refer to \cite{zhang:jstsp:14}.
\end{remark}
}}


\begin{remark}  \normalfont
(Common rate portion of RS)
Given that the code rate of the common stream $s_c$ is determined as $R_{c}\left(\mathbf{F},\mathbf{V}\right) = \min_{k \in \CMcal{K}_t^{{\text{int}}}, u \in \CMcal{K}_s} \big\{R_{c,k}^{\textnormal{bs}}\left(\mathbf{F},\mathbf{V}\right), R_{c,u}^{\textnormal{sat}}\left(\mathbf{F}\right)\big\}$, 
it is also possible to adjust each user's common rate portion. The total spectral efficiency achieved by SU $u$ is determined as $C_u + R_{p, u}^{\textnormal{sat}}\left(\bf{F}\right)$, where $C_u$ is the common rate portion of SU $u$ and $R_{p, u}^{\textnormal{sat}}\left(\bf{F}\right)$ is the spectral efficiency of private message for SU $u$. 
To ensure the decodability, we have $R_{c}\left(\mathbf{F},\mathbf{V}\right) \geq \sum_{u=1}^{K_s} C_u$.
Satisfying this decodability condition with equality, 
the sum spectral efficiency of the SUs is expressed as $ \sum_{u = 1}^{K_s} \{ C_u + R_{p, u}^{\textnormal{sat}}\left(\bf{F}\right) \} = R_{c}\left(\mathbf{F},\mathbf{V}\right) + \sum_{u = 1}^{K_s} R_{p, u}^{\textnormal{sat}}\left(\bf{F}\right)$.
As shown, an individual value of $C_u$ does not affect the sum spectral efficiency, thus we omit the design of the common rate portion $C_u$ in $\mathscr{P}_1$. 
On the contrary, if we consider the minimum spectral efficiency maximization problem, the common rate portion $C_u$ can affect the objective function so that it is required to carefully design $C_u$ \cite{kim:wcl:23}. 
\end{remark}



\begin{remark} \normalfont 
(Generalization to multi-layer RS) \label{remark:general}
For the sake of conciseness and brevity, we only consider 1-layer RS in this paper, which composes one type of common stream $s_{c}$ designed to be decodable for the users in $\{\CMcal{K}_s,\CMcal{K}_t^{{\text{int}}}\}$. 
It is of importance to note that the proposed method also can be extended to a generalized multi-layer RS setup. 
Using a multi-layer RS strategy, the sum spectral efficiency is further enhanced compared to that of a 1-layer RS strategy at the expense of increased decoding complexity.
In the multi-layer RS strategy, we can compose two types of common streams, where we call the super common stream $s_{sc}^{\textnormal{sat}}$ and the common stream $s_{c}^{\textnormal{sat}}$ for our convenience. 
The super common stream $s_{sc}^{\textnormal{sat}}$ is designed to be decodable in the users $\{\CMcal{K}_s,\CMcal{K}_t^{{\text{int}}}\}$, while the common stream $s_{c}^{\textnormal{sat}}$ is decodable in the users $\CMcal{K}_s$. 
By doing this, the super common stream $s_{sc}^{\textnormal{sat}}$ mainly helps the TUs to mitigate the ICI. To ensure that the super common stream is decodable in $\{\CMcal{K}_s,\CMcal{K}_t^{{\text{int}}}\}$, 
its spectral efficiency is determined by $R_{sc}^{\textnormal{sat}}\left(\mathbf{F},\mathbf{V}\right) = \min_{k \in \CMcal{K}_t^{{\text{int}}}, u \in \CMcal{K}_s} \{R_{sc,k}'\left(\mathbf{F},\mathbf{V}\right), R_{sc,u}'\left(\mathbf{F}\right)\}$. 
Additionally, the common stream $s_{c}^{\textnormal{sat}}$ mainly alleviates the IUI among the SUs, and its spectral efficiency is determined by $R_{c}^{\textnormal{sat}}\left(\mathbf{F}\right) = \min_{u \in \CMcal{K}_s} \{R_{c,u}'\left(\mathbf{F}\right)\}$ to ensure that the SUs can decode $s_{c}^{\textnormal{sat}}$. 
By doing this, we efficiently mitigate the IUI among the SUs as well as the ICI for the TUs, leading to the performance improvement. 
With the multi-layer RS strategy, the transmit signal of the LEO satellite is $\mathbf{x}_{\text{m}}^{\textnormal{sat}} = \mathbf{f}_{sc} s_{sc} + \mathbf{f}_{c} s_{c} + \sum_{i=1}^{K_s} \mathbf{f}_{p,i} s_{p,i}$. 
Even though we did not present the multi-layer RS in details due to the space limitation, we investigate the spectral efficiency performance of the multi-layer RS in Section \uppercase\expandafter{\romannumeral5}. 
\end{remark}


\section{Reformulation to Distributed Problem}
In this section, we propose a novel distributed precoding design in which CSIT sharing between the LEO satellite and terrestrial BS is not necessary.

\subsection{Representation to a Tractable Form}

To make $\mathscr{P}_1$ into a distributed form, it is of importance to reformulate the problem to a more tractable form. 
To this end, we first stack all the precoding matrices $\mathbf{F}$ and $\mathbf{V}$ and rewrite them as higher-dimensional precoding vectors $\Bar{\mathbf{f}} \in \mathbb{C}^{M(K_s+1)}$ and $\Bar{\mathbf{v}} \in \mathbb{C}^{NK_t}$, denoted as,
\begin{align}
    &\Bar{\mathbf{f}} = \left[\mathbf{f}_c^{\sf T}, \mathbf{f}_{p,1}^{\sf T}, \cdots, \mathbf{f}_{p,K_s}^{\sf T}\right]^{\sf T}, \;
    \Bar{\mathbf{v}} = \left[\mathbf{v}_{1}^{\sf T}, \cdots, \mathbf{v}_{K_t}^{\sf T}\right]^{\sf T}.\label{f_v_stack}
\end{align}
We define a unit vector whose the $k$-th element is 1 and the rest of the elements are zero as $\mathbf{u}_{k} = \left[0,\cdots,1,\cdots,0\right]^{\sf T} \in \mathbb{R}^{K_t}$, and a unit vector whose the $(u+1)$-th element 1 and the rest of the elements zero as $\mathbf{w}_{u} = \left[0,0,\cdots,1,\cdots,0\right]^{\sf T} \in \mathbb{R}^{(K_s+1)}$.

With \eqref{f_v_stack}, the spectral efficiency for the common stream of TU $k$ \eqref{R_com_TU_lower} is expressed by
\begin{align}
    \Bar{R}_{c,k}^{\textnormal{bs}}\left(\Bar{\mathbf{f}},\Bar{\mathbf{v}}\right) &= \log_2\left( \frac{\Bar{\mathbf{v}}^{\sf H}\mathbf{U}_{c,k}^{\textnormal{bs}}\Bar{\mathbf{v}} + \bar{\mathbf{f}}^{\sf H} \big( \mathbf{S}_{c,k}^{\textnormal{bs}} + \mathbf{C}_{c,k}^{\textnormal{bs}} \big) \bar{\mathbf{f}}}{\Bar{\mathbf{v}}^{\sf H}\mathbf{U}_{c,k}^{\textnormal{bs}}\Bar{\mathbf{v}} + \bar{\mathbf{f}}^{\sf H}\mathbf{C}_{c,k}^{\textnormal{bs}}\bar{\mathbf{f}}}\right), \label{re_R_com_TU}
\end{align}
where $\mathbf{S}_{c,k}^{\textnormal{bs}} = \mathbf{w}_{0}\mathbf{w}_{0}^{\sf H} \otimes \hat{\mathbf{z}}_{k}\hat{\mathbf{z}}_{k}^{\sf H} \in \mathbb{C}^{M(K_s+1) \times M(K_s+1)}$, $\mathbf{U}_{c,k}^{\textnormal{bs}} = \mathbf{I}_{K_t} \otimes \big(\hat{\mathbf{h}}_{k}\hat{\mathbf{h}}_{k}^{\sf H} + \mathbf{\Phi}_{k}^{\textnormal{bs}}\big) + \frac{\sigma^2}{P_s}\mathbf{I}_{NK_t} \in \mathbb{C}^{NK_t \times NK_t}$, and $\mathbf{C}_{c,k}^{\textnormal{bs}} = \mathbf{I}_{K_s+1} \otimes \big(\hat{\mathbf{z}}_{k}\hat{\mathbf{z}}_{k}^{\sf H} + \mathbf{\Phi}_{k}^{\textnormal{sat}}\big) - \mathbf{S}_{c,k}^{\textnormal{bs}} \in \mathbb{C}^{M(K_s+1) \times M(K_s+1)}$.
We also write the spectral efficiency for the common stream of SU $u$ \eqref{R_com_SU_lower} as
\begin{align}
    \Bar{R}_{c,u}^{\textnormal{sat}}\left(\Bar{\mathbf{f}}\right) = \log_2\left(\frac{\Bar{\mathbf{f}}^{\sf H} \big( \mathbf{S}_{c,u}^{\textnormal{sat}} + \mathbf{U}_{c,u}^{\textnormal{sat}} \big)\Bar{\mathbf{f}}}{\Bar{\mathbf{f}}^{\sf H}\mathbf{U}_{c,u}^{\textnormal{sat}}\Bar{\mathbf{f}}} \right),\label{re_R_com_SU}
\end{align}
where $\mathbf{S}_{c,k}^{\textnormal{sat}} = \mathbf{w}_{0}\mathbf{w}_{0}^{\sf H} \otimes \hat{\mathbf{g}}_{u}\hat{\mathbf{g}}_{u}^{\sf H} \in \mathbb{C}^{M(K_s+1) \times M(K_s+1)}$, $\mathbf{U}_{c,k}^{\textnormal{sat}} = \mathbf{I}_{K_s+1} \otimes \big(\hat{\mathbf{g}}_{u}\hat{\mathbf{g}}_{u}^{\sf H} + \mathbf{\Psi}_{u}^{\textnormal{sat}}\big) + \frac{\sigma^2}{P_s}\mathbf{I}_{M(K_s+1)} - \mathbf{S}_{c,u}^{\textnormal{sat}} \in \mathbb{C}^{M(K_s+1) \times M(K_s+1)}$.
Similar to this, the spectral efficiency for the private stream of TU $k$ \eqref{R_pri_TU_lower} is also written by
\begin{align}
    \Bar{R}_{p,k}^{\textnormal{bs}}\left(\Bar{\mathbf{f}},\Bar{\mathbf{v}}\right) &= \log_2\left( \frac{ \Bar{\mathbf{v}}^{\sf H} \big( \mathbf{S}_{p,k}^{\textnormal{bs}} + \mathbf{U}_{p,k}^{\textnormal{bs}} \big) \Bar{\mathbf{v}} + \Bar{\mathbf{f}}^{\sf H}\mathbf{C}_{p,k}^{\textnormal{bs}}\Bar{\mathbf{f}} }{\Bar{\mathbf{v}}^{\sf H}\mathbf{U}_{p,k}^{\textnormal{bs}}\Bar{\mathbf{v}} + \Bar{\mathbf{f}}^{\sf H}\mathbf{C}_{p,k}^{\textnormal{bs}}\Bar{\mathbf{f}} }\right),
    \label{re_R_pri_TU}
\end{align}
where $\mathbf{S}_{p,k}^{\textnormal{bs}} = \mathbf{u}_{k}\mathbf{u}_{k}^{\sf H} \otimes \hat{\mathbf{h}}_{k}\hat{\mathbf{h}}_{k}^{\sf H} \in \mathbb{C}^{NK_t \times NK_t}$, $\mathbf{U}_{p,k}^{\textnormal{bs}} = \mathbf{I}_{K_t} \otimes \big(\hat{\mathbf{h}}_{k}\hat{\mathbf{h}}_{k}^{\sf H} + \mathbf{\Phi}_{k}^{\textnormal{bs}}\big) + \frac{\sigma^2}{P_t}\mathbf{I}_{NK_t} - \mathbf{S}_{p,k}^{\textnormal{bs}} \in \mathbb{C}^{NK_t \times NK_t}$, and $\mathbf{C}_{p,k}^{\textnormal{bs}} = \mathbf{I}_{K_s+1} \otimes \frac{P_s}{P_t}\big(\hat{\mathbf{z}}_{k}\hat{\mathbf{z}}_{k}^{\sf H} + \mathbf{\Phi}_{k}^{\textnormal{sat}}\big) - \mathbf{w}_{0}\mathbf{w}_{0}^{\sf H} \otimes \big(\hat{\mathbf{z}}_{k}\hat{\mathbf{z}}_{k}^{\sf H} + \mathbf{\Phi}_{k}^{\textnormal{sat}}\big) \in \mathbb{C}^{M(K_s+1) \times M(K_s+1)}$.
Next, we present the spectral efficiency for the private stream of SU $u$ \eqref{R_pri_SU_lower} as
\begin{align}
    \Bar{R}_{p,u}^{\textnormal{sat}}\left(\Bar{\mathbf{f}}\right) = \log_2\left( \frac{\Bar{\mathbf{f}}^{\sf H} \big( \mathbf{S}_{p,u}^{\textnormal{sat}} + \mathbf{U}_{p,u}^{\textnormal{sat}} \big) \Bar{\mathbf{f}}}{\Bar{\mathbf{f}}^{\sf H}\mathbf{U}_{p,u}^{\textnormal{sat}}\Bar{\mathbf{f}}} \right),\label{re_R_pri_SU}
\end{align}
where $\mathbf{S}_{p,u}^{\textnormal{sat}} = \mathbf{w}_{u}\mathbf{w}_{u}^{\sf H} \otimes \hat{\mathbf{g}}_{u}\hat{\mathbf{g}}_{u}^{\sf H}  \in \mathbb{C}^{M(K_s+1) \times M(K_s+1)}$ and $\mathbf{U}_{p,u}^{\textnormal{sat}} = \textnormal{Blkd}\left[0, \mathbf{I}_{K_s}\right] \otimes \big(\hat{\mathbf{g}}_{u}\hat{\mathbf{g}}_{u}^{\sf H} + \mathbf{\Psi}_{u}^{\textnormal{sat}}\big) + \frac{\sigma^2}{P_s}\textnormal{Blkd}\left[\mathbf{0}_{M},\mathbf{I}_{MK_s}\right] - \mathbf{S}_{p,u}^{\textnormal{sat}} \in \mathbb{C}^{M(K_s+1) \times M(K_s+1)}$.
Here, $\mathbf{0}_{M} \in\mathbb{C}^{M \times M} $ is the null matrix and $\textnormal{Blkd}\left[\mathbf{0}_{M},\mathbf{I}_{MK_s}\right] \in \mathbb{C}^{M(K_s+1) \times M(K_s+1)}$ is a block-diagonal matrix concatenating $\mathbf{0}_{M},\mathbf{I}_{MK_s}$.
In \eqref{re_R_com_TU}, \eqref{re_R_com_SU}, \eqref{re_R_pri_TU}, and \eqref{re_R_pri_SU}, we assume that the higher-dimensional precoding vectors $\bar{\mathbf{f}}$ and $\Bar{\mathbf{v}}$ satisfy the transmit power constraint with equality, i.e., $\lVert\bar{\mathbf{f}}\rVert^2 = 1$ and $\lVert\bar{\mathbf{v}}\rVert^2 = 1$. Note that this assumption does not affect the optimality \cite{park:twc:23}. 
Using our representation, the sum spectral efficiency in \eqref{opt_prob} is rewritten as 
\begin{align}
    \Bar{R}_{\textnormal{sum}}\left(\bar{\mathbf{f}},\bar{\mathbf{v}}\right) &= 
    {\color{black}{
    \begin{bmatrix}
        \min_{k \in \CMcal{K}_t^{{\text{int}}}, u \in \CMcal{K}_s} \left\{\Bar{R}_{c,k}^{\textnormal{bs}}\left(\Bar{\mathbf{f}},\Bar{\mathbf{v}}\right), \Bar{R}_{c,u}^{\textnormal{sat}}\left(\Bar{\mathbf{f}}\right)\right\}\\
        + \sum_{i=1}^{K_s} \Bar{R}_{p,i}^{\textnormal{sat}}\left(\Bar{\mathbf{f}}\right) + \sum_{j=1}^{K_t} \Bar{R}_{p,j}^{\textnormal{bs}}\left(\Bar{\mathbf{f}},\Bar{\mathbf{v}}\right)
    \end{bmatrix}
    }}.\label{R_sum}
\end{align}
Despite this, however, the sum spectral efficiency \eqref{R_sum} still cannot be optimized in a distributed manner because the LEO satellite precoding vector $\bar {\bf{f}}$ and the terrestrial BS precoding vector $\bar {\bf{v}}$ are entangled in ${\Bar{R}_{c,k}^{\textnormal{bs}}\left(\Bar{\mathbf{f}},\Bar{\mathbf{v}}\right)}$ and  ${\Bar{R}_{p,k}^{\textnormal{bs}}\left(\Bar{\mathbf{f}},\Bar{\mathbf{v}}\right)}$. 
In the following subsection, we explain how to decouple  ${\Bar{R}_{c,k}^{\textnormal{bs}}\left(\Bar{\mathbf{f}},\Bar{\mathbf{v}}\right)}$ and ${\Bar{R}_{p,k}^{\textnormal{bs}}\left(\Bar{\mathbf{f}},\Bar{\mathbf{v}}\right)}$. 

\subsection{Transformation to Distributed Forms}
{\textbf{Private stream spectral efficiency decoupling}}: We first decouple ${\Bar{R}_{p,k}^{\textnormal{bs}}\left(\Bar{\mathbf{f}},\Bar{\mathbf{v}}\right)}$ defined in \eqref{re_R_pri_TU}. A key challenge in decoupling of ${\Bar{R}_{p,k}^{\textnormal{bs}}\left(\Bar{\mathbf{f}},\Bar{\mathbf{v}}\right)}$ is that the IUI term $\Bar{\mathbf{v}}^{\sf H}\mathbf{U}_{p,k}^{\textnormal{bs}}\Bar{\mathbf{v}}$, a function of $\bar {\bf{v}}$, and the ICI term $\Bar{\mathbf{f}}^{\sf H}\mathbf{C}_{p,k}^{\textnormal{bs}}\Bar{\mathbf{f}}$, a function of $\bar {\bf{f}}$, 
coexist in the numerator and the denominator as shown in \eqref{re_R_pri_TU}. 
To resolve this, we consider the ergodic spectral efficiency where the expectation is taken over the randomness associated with the imperfect knowledge of the channel fading process: $\mathbb{E}_{\{\hat {\bf{h}}_k, \hat {\bf{z}}_k\}} \big[ \Bar{R}_{p,k}^{\textnormal{bs}}\left(\Bar{\mathbf{f}},\Bar{\mathbf{v}}\right) \big]$. For simplicity, we drop the notation of $\left\{\hat {\bf{h}}_k, \hat {\bf{z}}_k\right\}$ from the expectation. Then we have 
\begin{align}
    \mathbb{E} \left[ \Bar{R}_{p,k}^{\textnormal{bs}}\left(\Bar{\mathbf{f}},\Bar{\mathbf{v}}\right) \right]  
    &\overset{(a)}{\approx} \mathbb{E} \left[ \log_2\left(\frac{\Bar{\mathbf{v}}^{\sf H}\mathbf{S}_{p,k}^{\textnormal{bs}}\Bar{\mathbf{v}}}{\Bar{\mathbf{v}}^{\sf H}\mathbf{U}_{p,k}^{\textnormal{bs}}\Bar{\mathbf{v}} + \Bar{\mathbf{f}}^{\sf H}\mathbf{C}_{p,k}^{\textnormal{bs}}\Bar{\mathbf{f}}}\right) \right] \nonumber\\
    &\overset{(b)}{\approx} \mathbb{E} \left[ \log_2\left(\frac{\Bar{\mathbf{v}}^{\sf H}\mathbf{S}_{p,k}^{\textnormal{bs}}\Bar{\mathbf{v}}}{ \mathbb{E}\big[\Bar{\mathbf{v}}^{\sf H}\mathbf{U}_{p,k}^{\textnormal{bs}}\Bar{\mathbf{v}}\big] + \Bar{\mathbf{f}}^{\sf H}\mathbf{C}_{p,k}^{\textnormal{bs}}\Bar{\mathbf{f}}}\right) \right] \nonumber\\
    &\overset{(c)}{=} \mathbb{E}\left[ \log_2\left(\frac{\Bar{\mathbf{v}}^{\sf H}\mathbf{S}_{p,k}^{\textnormal{bs}}\Bar{\mathbf{v}}}{ \Bar{\mathbf{f}}^{\sf H}\big( \mathbf{C}_{p,k}^{\textnormal{bs}} + \epsilon_k\mathbf{I}_{M(K_s+1)} \big) \Bar{\mathbf{f}}}\right)\right]
    ,\label{re_R_pri_TU_approx}    
\end{align}
where $(a)$ comes from the high SNR assumption and $(b)$ follows Jensen's inequality. 
As observed in $(c)$, by averaging the IUI term, it is captured as a constant, defined as $\mathbb{E} \big[ \Bar{\mathbf{v}}^{\sf H}\mathbf{U}_{p,k}^{\textnormal{bs}}\Bar{\mathbf{v}} \big] = \epsilon_k$. 
This allows us to disentangle the $\bar {\bf{f}}$-related term and the $\bar {\bf{v}}$-related term into the numerator and the denominator in \eqref{re_R_pri_TU_approx}.
With \eqref{re_R_pri_TU_approx}, by rearranging the ergodic sum spectral efficiency of the private streams, we obtain \eqref{eq:R_private_dist} at the top of the next page.
\begin{figure*}
    \begin{align}
    &\sum_{i=1}^{K_s} \mathbb{E} \left[\Bar{R}_{p,i}^{\textnormal{sat}}\left(\Bar{\mathbf{f}}\right)\right] + \sum_{j=1}^{K_t} \mathbb{E}\left[\Bar{R}_{p,j}^{\textnormal{bs}}\left(\Bar{\mathbf{f}},\Bar{\mathbf{v}}\right)\right]
    \mathop{\ge}^{(a)} \sum_{i=1}^{K_s} \mathbb{E}\left[ \log_2\left( \frac{\Bar{\mathbf{f}}^{\sf H} \left( \mathbf{S}_{p,i}^{\textnormal{sat}} + \mathbf{U}_{p,i}^{\textnormal{sat}} \right) \Bar{\mathbf{f}}}{\Bar{\mathbf{f}}^{\sf H}\mathbf{U}_{p,i}^{\textnormal{sat}}\Bar{\mathbf{f}}} \right)\right] + \sum_{j=1}^{K_t} \mathbb{E}\left[ \log_2\left(\frac{\Bar{\mathbf{v}}^{\sf H}\mathbf{S}_{p,j}^{\textnormal{bs}}\Bar{\mathbf{v}}}{ \Bar{\mathbf{f}}^{\sf H}\left( \mathbf{C}_{p,j}^{\textnormal{bs}} + \epsilon_j \mathbf{I}_{M(K_s+1)} \right) \Bar{\mathbf{f}}}\right) \right] \nonumber\\
    &= \mathbb{E}\left[ \log_2\left( \prod_{i=1}^{K_s} \frac{\Bar{\mathbf{f}}^{\sf H} \big( \mathbf{S}_{p,i}^{\textnormal{sat}} + \mathbf{U}_{p,i}^{\textnormal{sat}} \big) \Bar{\mathbf{f}}}{\Bar{\mathbf{f}}^{\sf H}\mathbf{U}_{p,i}^{\textnormal{sat}}\Bar{\mathbf{f}}} \prod_{j=1}^{K_t} \frac{\Bar{\mathbf{v}}^{\sf H}\mathbf{S}_{p,j}^{\textnormal{bs}}\Bar{\mathbf{v}}}{ \Bar{\mathbf{f}}^{\sf H} \big( \mathbf{C}_{p,j}^{\textnormal{bs}} + \epsilon_j\mathbf{I}_{M(K_s+1)} \big) \Bar{\mathbf{f}} } \right) \right] 
    \overset{(b)}{=} \mathbb{E}\left[ \log_2\left( \frac{ \prod_{i=1}^{K_s} \Bar{\mathbf{f}}^{\sf H} \big( \mathbf{S}_{p,i}^{\textnormal{sat}} + \mathbf{U}_{p,i}^{\textnormal{sat}} \big) \Bar{\mathbf{f}}}{ \prod_{i=1}^{K_s} \Bar{\mathbf{f}}^{\sf H}\mathbf{U}_{p,i}^{\textnormal{sat}}\Bar{\mathbf{f}} \prod_{j=1}^{K_t} \Bar{\mathbf{f}}^{\sf H}\big( \mathbf{C}_{p,j}^{\textnormal{bs}} + \epsilon_j \mathbf{I}_{M(K_s+1)} \big) \Bar{\mathbf{f}}} \prod_{j=1}^{K_t} \Bar{\mathbf{v}}^{\sf H}\mathbf{S}_{p,j}^{\textnormal{bs}}\Bar{\mathbf{v}} \right) \right] \nonumber\\
    &= \sum_{i=1}^{K_s} \mathbb{E}\left[ \log_2\left[ \frac{\Bar{\mathbf{f}}^{\sf H} \big( \mathbf{S}_{p,i}^{\textnormal{sat}} + \mathbf{U}_{p,i}^{\textnormal{sat}} \big) \Bar{\mathbf{f}}}{\Bar{\mathbf{f}}^{\sf H}\mathbf{U}_{p,i}^{\textnormal{sat}}\Bar{\mathbf{f}} \cdot 
    \big\{\prod_{j=1}^{K_t}\Bar{\mathbf{f}}^{\sf H}\big( \mathbf{C}_{p,j}^{\textnormal{bs}} + \epsilon_j \mathbf{I}_{M(K_s+1)} \big)\Bar{\mathbf{f}} \big\}^{\frac{1}{K_s}}} \right] \right] + \underbrace{\sum_{j=1}^{K_t} \mathbb{E}\left[ \log_2\left( \Bar{\mathbf{v}}^{\sf H}\mathbf{S}_{p,j}^{\textnormal{bs}}\Bar{\mathbf{v}} \right)\right] }_{(c)}, \label{eq:R_private_dist}
\end{align}
\hrule
\end{figure*}
In \eqref{eq:R_private_dist}, $(a)$ follows \eqref{re_R_pri_TU_approx}. 
Additionally, the ICI term $ \Bar{\mathbf{f}}^{\sf H}\big( \mathbf{C}_{p,k}^{\textnormal{bs}} + \epsilon_k\mathbf{I}_{M(K_s+1)} \big) \Bar{\mathbf{f}}$, originally placed in $\Bar{R}_{p,k}^{\textnormal{bs}}\left(\Bar{\mathbf{f}},\Bar{\mathbf{v}}\right)$, comes out as a product of the leakage interference $\big\{\prod_{j=1}^{K_t}\Bar{\mathbf{f}}^{\sf H}\big( \mathbf{C}_{p,j}^{\textnormal{bs}} + \epsilon_j \mathbf{I}_{M(K_s+1)} \big)\Bar{\mathbf{f}} \big\}^{\frac{1}{K_s}}$, which is a function of only $\bar {\bf{f}}$. 
Thanks to this, we integrate this leakage interference term into $\bar R_{p,i} (\bar {\bf{f}})$ as shown in $(b)$. 
In \eqref{eq:R_private_dist}, the private stream spectral efficiency is reshaped into two decoupled terms, wherein the first term is a function of $\bar {\bf{f}}$ and the second term is a function of $\bar {\bf{v}}$. This enables to tackle the problem in a distributed fashion. 
Nonetheless, we observe in $(c)$ that the IUI term vanishes since we take average and pull out from $\Bar{R}_{p,k}^{\textnormal{bs}}\left(\Bar{\mathbf{f}},\Bar{\mathbf{v}}\right)$ for decoupling \eqref{re_R_pri_TU_approx}. 
To compensate this, we restore the vanished IUI term in $(c)$ as follows: 
\begin{align} \label{eq:restore}
    \sum_{j=1}^{K_t} \mathbb{E}\left[ \log_2\left( \Bar{\mathbf{v}}^{\sf H}\mathbf{S}_{p,j}^{\textnormal{bs}}\Bar{\mathbf{v}} \right)\right] 
    &= \mathbb{E} \left[ \log_2\left( \prod_{j=1}^{K_t} \frac{\Bar{\mathbf{v}}^{\sf H}\mathbf{S}_{p,j}^{\textnormal{bs}}\Bar{\mathbf{v}} \cdot \Bar{\mathbf{v}}^{\sf H}\mathbf{U}_{p,j}^{\textnormal{bs}}\Bar{\mathbf{v}} }{ \Bar{\mathbf{v}}^{\sf H}\mathbf{U}_{p,j}^{\textnormal{bs}}\Bar{\mathbf{v}} } \right) \right] \nonumber\\
    &= \mathbb{E} \left[ \log_2\left( \prod_{j=1}^{K_t} \frac{\Bar{\mathbf{v}}^{\sf H}\mathbf{S}_{p,j}^{\textnormal{bs}}\Bar{\mathbf{v}} }{ \Bar{\mathbf{v}}^{\sf H}\mathbf{U}_{p,j}^{\textnormal{bs}}\Bar{\mathbf{v}} } \right) \right] + \hat{\epsilon},
\end{align}
where $\hat{\epsilon} =  \mathbb{E} \big[ \sum_{j=1}^{K_t}\log_2\big( \Bar{\mathbf{v}}^{\sf H}\mathbf{U}_{p,j}^{\textnormal{bs}}\Bar{\mathbf{v}}\big) \big]$. 
Inserting $\hat \epsilon$ into the $\bar {\bf{f}}$-related term, we get 
\begin{align}
    &\sum_{i=1}^{K_s} \mathbb{E}\left[\Bar{R}_{p,i}^{\textnormal{sat}}\left(\Bar{\mathbf{f}}\right)\right] + \sum_{j=1}^{K_t} \mathbb{E}\left[\Bar{R}_{p,j}^{\textnormal{bs}}\left(\Bar{\mathbf{f}},\Bar{\mathbf{v}}\right)\right] \nonumber\\
    &\gtrsim \sum_{i=1}^{K_s} \mathbb{E}\left[\Bar{\Gamma}_{p,i}^{\textnormal{sat}}\left(\bar{\mathbf{f}}\right)\right] + \sum_{j=1}^{K_t} \mathbb{E}\left[\Bar{\Gamma}_{p,j}^{\textnormal{bs}}\left(\bar{\mathbf{v}}\right)\right],\label{R_sum_2}
\end{align}
where   
\begin{align}
    &\mathbb{E}\left[ \Bar{\Gamma}_{p,u}^{\textnormal{sat}}\left(\bar{\mathbf{f}}\right) \right] \nonumber\\
    &= \mathbb{E}\left[ \log_2\left[ \frac{ 2^{\frac{\hat{\epsilon}}{K_s}} \cdot \Bar{\mathbf{f}}^{\sf H} \left( \mathbf{S}_{p,u}^{\textnormal{sat}} + \mathbf{U}_{p,u}^{\textnormal{sat}}  \right) \Bar{\mathbf{f}}}{\Bar{\mathbf{f}}^{\sf H}\mathbf{U}_{p,u}^{\textnormal{sat}}\Bar{\mathbf{f}} \cdot \left\{\prod_{j=1}^{K_t}\Bar{\mathbf{f}}^{\sf H}\left( \mathbf{C}_{p,j}^{\textnormal{bs}} + \epsilon_j \mathbf{I}_{M(K_s+1)} \right)\Bar{\mathbf{f}} \right\}^{\frac{1}{K_s}}} \right]\right], \label{er_R_p_u}\\   
    &\mathbb{E}\left[ \Bar{\Gamma}_{p,k}^{\textnormal{bs}}\left(\bar{\mathbf{v}}\right) \right] 
    = \mathbb{E}\left[ \log_2\left( \frac{\Bar{\mathbf{v}}^{\sf H}\mathbf{S}_{p,k}^{\textnormal{bs}}\Bar{\mathbf{v}} }{ \Bar{\mathbf{v}}^{\sf H}\mathbf{U}_{p,k}^{\textnormal{bs}}\Bar{\mathbf{v}} } \right)\right].\label{er_R_p_k}
\end{align}
One could be confused regarding the processes in \eqref{eq:R_private_dist} and \eqref{eq:restore} because we pull out the averaged IUI term from $\Bar{R}_{p,k}^{\textnormal{bs}}\left(\Bar{\mathbf{f}},\Bar{\mathbf{v}}\right)$ then restore it again. The rationale is as follows. 
As shown in \eqref{re_R_pri_TU_approx}, $\Bar{R}_{p,k}^{\textnormal{bs}}\left(\Bar{\mathbf{f}},\Bar{\mathbf{v}}\right)$ entails the $\bar {\bf{v}}$-related IUI term and the $\bar {\bf{f}}$-related ICI term. To decouple this, it is required to detach the ICI term; yet it is infeasible due to the $\bar {\bf{v}}$ dependence. To resolve this, we take average to the IUI term, effectively eliminating the dependence of $\bar {\bf{v}}$. After decoupling, however, $\Bar{R}_{p,k}^{\textnormal{bs}}\left(\Bar{\mathbf{f}},\Bar{\mathbf{v}}\right)$ loses the IUI term as observed in (c) of \eqref{eq:R_private_dist}. This brings severe performance degradation in $\Bar{R}_{p,k}^{\textnormal{bs}}\left(\Bar{\mathbf{f}},\Bar{\mathbf{v}}\right)$ since we cannot take into account the IUI in designing $\bar {\bf{v}}$. To address this, we restore the IUI term $\Bar{\mathbf{v}}^{\sf H}\mathbf{U}_{p,j}^{\textnormal{bs}}\Bar{\mathbf{v}}$ into $\Bar{R}_{p,k}^{\textnormal{bs}}\left(\Bar{\mathbf{f}},\Bar{\mathbf{v}}\right)$ as in \eqref{eq:restore}.
Upon this, it is rendered that $\Bar{\Gamma}_{p,u}^{\textnormal{sat}}\left(\bar{\mathbf{f}}\right)$ and $\Bar{\Gamma}_{p,k}^{\textnormal{bs}}\left(\bar{\mathbf{v}}\right)$ in \eqref{er_R_p_u} and \eqref{er_R_p_k} 
are functions of only $\bar {\bf{f}}$ and $\bar {\bf{v}}$ respectively, thus each can be maximized in a distributed fashion without CSIT sharing. 
Note that the satellite is required to have $\epsilon_k$ and $\hat \epsilon$ to maximize $\Bar{\Gamma}_{p,u}^{\textnormal{sat}}\left(\bar{\mathbf{f}}\right)$. 
Fortunately, $\epsilon_k$ and $\hat \epsilon$ are constants that does not vary over channels. 
For this reason, it is easy to deliver these values to the LEO satellite, for instance by embedding into the control channels without incurring much overheads.

{\textbf{Common stream spectral efficiency decoupling}}:
Subsequently, we reform the common stream spectral efficiency $\bar R_{c, k}^{\textnormal{bs}} (\bar {\bf{f}}, \bar {\bf{v}})$ \eqref{re_R_com_TU} in a distributed manner. 
A critical obstacle to decouple $\bar R_{c, k}^{\textnormal{bs}} (\bar {\bf{f}}, \bar {\bf{v}})$ is that $\Bar{\mathbf{v}}^{\sf H}\mathbf{U}_{c,k}^{\textnormal{bs}}\Bar{\mathbf{v}}$ exists in the denominator of $\bar R_{c, k}^{\textnormal{bs}} (\bar {\bf{f}}, \bar {\bf{v}})$. 
To resolve this, we express the ergodic spectral efficiency of the common stream for TU $k$ as 
\begin{align}
    \mathbb{E} \left[ \Bar{R}_{c,k}^{\textnormal{bs}}\left(\Bar{\mathbf{f}},\Bar{\mathbf{v}}\right) \right] 
    &= \mathbb{E} \left[ \log_2\left( 1 + \frac{ \bar{\mathbf{f}}^{\sf H}  \mathbf{S}_{c,k}^{\textnormal{bs}} \bar{\mathbf{f}}}{\Bar{\mathbf{v}}^{\sf H}\mathbf{U}_{c,k}^{\textnormal{bs}}\Bar{\mathbf{v}} + \bar{\mathbf{f}}^{\sf H}\mathbf{C}_{c,k}^{\textnormal{bs}}\bar{\mathbf{f}}}\right) \right] \nonumber\\
    &\overset{(a)}{\geq} 
    \mathbb{E} \left[ \log_2\left( 1 + \frac{ \bar{\mathbf{f}}^{\sf H}  \mathbf{S}_{c,k}^{\textnormal{bs}} \bar{\mathbf{f}}}{ \mathbb{E} \big[ \Bar{\mathbf{v}}^{\sf H}\mathbf{U}_{c,k}^{\textnormal{bs}}\Bar{\mathbf{v}} \big]  + \bar{\mathbf{f}}^{\sf H}\mathbf{C}_{c,k}^{\textnormal{bs}}\bar{\mathbf{f}}}\right) \right],\label{er_R_c}
\end{align}
where the expectation is taken over the randomness associated with the imperfect knowledge of the channel fading process and $(a)$ follows Jensen's inequality.
{\color{black}{Notice that with distributed precoding, the satellite precoding vector $\bar {\bf{f}}$ and the terrestrial BS precoding vector $\bar {\bf{v}}$ are separately devised, so that the independence between the desired signal power and the interference power holds. }}
In \eqref{er_R_c}, we make the IUI term as a constant by averaging it, defined as $\mathbb{E} \big[ \Bar{\mathbf{v}}^{\sf H}\mathbf{U}_{c,k}^{\textnormal{bs}}\Bar{\mathbf{v}} \big] = \omega_k$, so that we transform $\Bar{R}_{c,k}^{\textnormal{bs}}\left(\Bar{\mathbf{f}},\Bar{\mathbf{v}}\right)$, which was originally a joint function of $\bar{\mathbf{f}}$ and $\bar{\mathbf{v}}$, into 
\begin{align}
    \mathbb{E} \left[ \Bar{R}_{c,k}^{\textnormal{bs}}\left(\Bar{\mathbf{f}},\Bar{\mathbf{v}}\right) \right] 
    &\geq \mathbb{E} \left[ \Bar{\Gamma}_{c,k}^{\textnormal{bs}}\left(\bar{\mathbf{f}}\right) \right] \nonumber\\
    &= \mathbb{E} \left[ \log_2\left( \frac{ \bar{\mathbf{f}}^{\sf H} \left( \mathbf{S}_{c,k}^{\textnormal{bs}} + \mathbf{C}_{c,k}^{\textnormal{bs}} + \omega_k \mathbf{I}_{M(K_s + 1)} \right) \bar{\mathbf{f}}}{ \bar{\mathbf{f}}^{\sf H} \left( \mathbf{C}_{c,k}^{\textnormal{bs}} + \omega_k \mathbf{I}_{M(K_s+1)} \right) \bar{\mathbf{f}}}\right) \right].\label{gamma_R_c_k}
\end{align}
By doing this, provided that $\omega_k$ is known to the satellite, it is feasible to maximize the common stream spectral efficiency at the satellite without sharing CSIT. 
Similar to $\epsilon_k$ and $\hat \epsilon$, $\omega_k$ is a constant, thus it is effortless to deliver $\omega_k$ to the satellite.

\begin{remark} \normalfont
(Comparison to other decoupling methods)
We clarify the key distinguishing aspects of our decoupling method as follows.
SLNR \cite{sadek:twc:07} considers the leakage terms instead of the correct IUI terms, so that it fails to capture the IUI term in a proper way. 
This not only causes performance degradation, but also makes it difficult to apply SLNR to the RS strategy. Since the IUI is not counted, it is infeasible to characterize a lower bound on the spectral efficiency of the common stream. 
SILNR \cite{han:tcom:21} decouples the spectral efficiency by forcibly setting ICI and IUI to zero. This way maybe useful when the amount of ICI and IUI are small, i.e., perfect CSIT; yet it is not appropriate when imperfect CSIT is taken into account and the amount of ICI and IUI are not negligible. 
Similar to SILNR, IUI-ICI separation \cite{Choi:twc:12} sets the ICI as zero to decouple the spectral efficiency and use the WMMSE method. As explained above, forcing the ICI to zero in the imperfect CSIT setup is not justified. 
Additionally, applying the IUI-ICI separation and the WMMSE developed in \cite{Choi:twc:12} with the RS strategy is not straightforward. 

In summary, our method stands out because \lowercase\expandafter{\romannumeral1}) the IUI term is properly captured, \lowercase\expandafter{\romannumeral2}) it is applicable in the imperfect CSIT setup, and \lowercase\expandafter{\romannumeral3}) it is compatible with the RS strategy. 
\end{remark}


{\textbf{Distributed problem formulation}}:
Up to this point, we transform $\mathbb{E} \big[\bar R_{c, k}^{\textnormal{bs}} (\bar {\bf{f}}, \bar {\bf{v}})\big]$ and $\mathbb{E}\big[\Bar{R}_{p,k}^{\textnormal{bs}}\left(\Bar{\mathbf{f}},\Bar{\mathbf{v}}\right)\big]$ to $\mathbb{E} \big[ \Bar{\Gamma}_{c,k}^{\textnormal{bs}}\left(\bar{\mathbf{f}}\right) \big]$ and $\mathbb{E}\big[\Bar{\Gamma}_{p,k}^{\textnormal{bs}}\left(\bar{\mathbf{v}}\right)\big]$, wherein distributed precoding optimization is feasible by using our spectral efficiency decoupling. 
Based on this, we reformulate our main problem $\mathscr{P}_1$ in \eqref{opt_prob}. 
We first recall that the expectation considered in the ergodic spectral efficiency \eqref{R_sum_2} and \eqref{gamma_R_c_k} is associated with the randomness of the imperfect knowledge on channel fading process, i.e., $\mathbb{E}_{\{\hat {\bf{h}}_k, {\hat {\bf{g}}}_u, \hat {\bf{z}}_k \}} [ \cdot ]$.  
Since the LEO satellite and the terrestrial BS are assumed to have the estimated CSIT $\hat {\bf{g}}_u, \hat {\bf{z}}_k$, (the satellite channel) and $\hat {{\bf{h}}}_k$ (the terrestrial channel), maximizing the ergodic spectral efficiency is equivalent to maximizing the spectral efficiency by using the estimated CSIT given per each channel block \cite{park:twc:23, joudeh:tcom:16}. 
For this reason, we can omit $\mathbb{E}\left[ \cdot \right]$ in \eqref{R_sum_2} and \eqref{gamma_R_c_k}. 
Consequently, we formulate the distributed precoding optimization problem as
\begin{align}
    \mathscr{P}_2: \; &\underset{\bar{\mathbf{f}},\bar{\mathbf{v}}}{\textnormal{maximize}}
    \;\; 
    {\color{black}{
    \begin{bmatrix}
        \min_{k \in \CMcal{K}_t^{{\text{int}}}, u \in \CMcal{K}_s} \left\{ \Bar{\Gamma}_{c,k}^{\textnormal{bs}} \left(\bar{\mathbf{f}}\right) , \Bar{R}_{c,u}^{\textnormal{sat}}\left(\Bar{\mathbf{f}}\right)\right\}\\
        + \sum_{i=1}^{K_s} \Bar{\Gamma}_{p,i}^{\textnormal{sat}}\left(\bar{\mathbf{f}}\right) + \sum_{j=1}^{K_t} \Bar{\Gamma}_{p,j}^{\textnormal{bs}} \left(\bar{\mathbf{v}}\right)
    \end{bmatrix}
    }}
    ,\label{re_opt_prob}\\
    &\textnormal{subject to} \;\,\, \lVert\bar{\mathbf{f}}\rVert^2 = 1,\;
    \lVert\bar{\mathbf{v}}\rVert^2 = 1.\label{st_stack_V}
\end{align}
It is noteworthy that in $\mathscr{P}_2$, all the spectral efficiencies in the objective function are exclusively determined by $\bar {\bf{f}}$ and $\bar {\bf{v}}$ respectively. 
This eliminates the necessity of CSIT sharing in solving $\mathscr{P}_2$.
Furthermore, the precoding vectors $\bar{\mathbf{f}},\bar{\mathbf{v}}$ respectively can be normalized by dividing the numerator and denominator of Rayleigh quotient with $\lVert\bar{\mathbf{f}}\rVert^2$ and $\lVert\bar{\mathbf{v}}\rVert^2$, without affecting the objective function. Thanks to this reason, the constraint \eqref{st_stack_V} can be omitted. 

\begin{figure}[t]
 \renewcommand{\figurename}{Fig.}
    \centering
    \includegraphics[width=9cm]{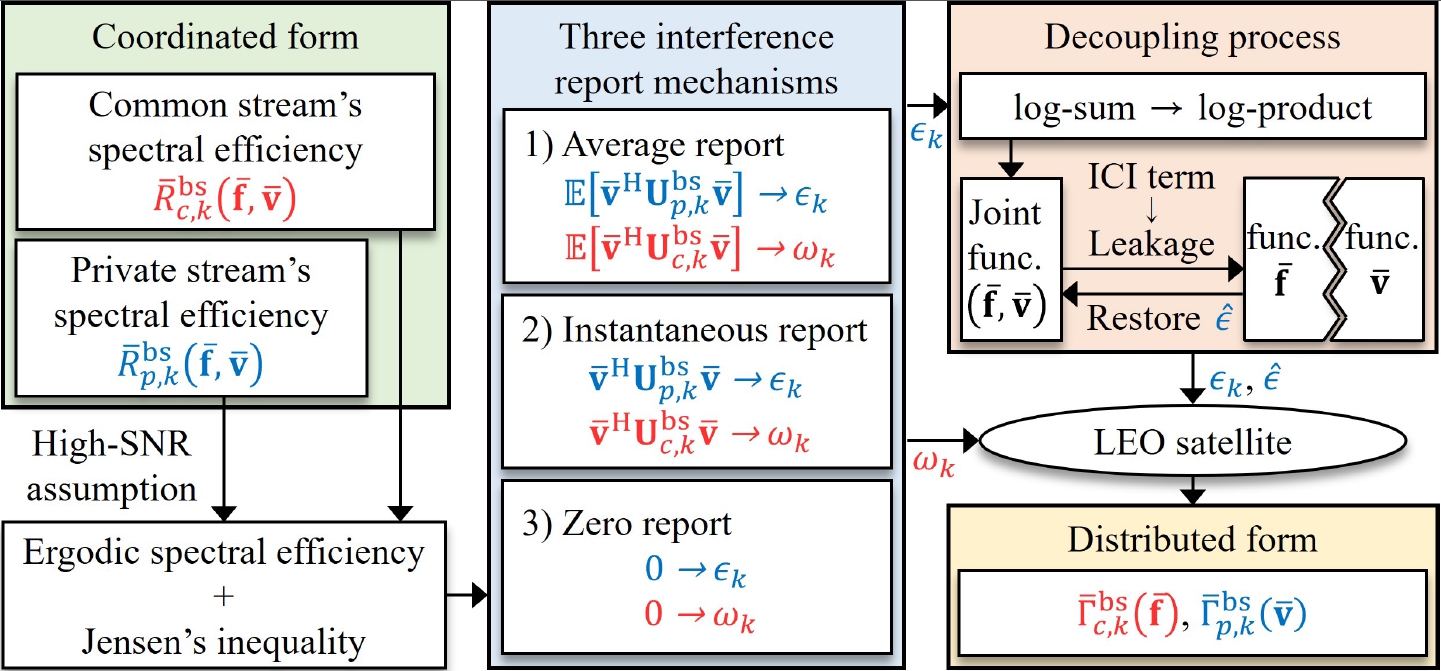}
    \caption{{\color{black}{The decoupling process from coordinated form to distributed form.}}}
    \label{decoupling process}
\end{figure}

{\color{black}{For ease of understanding, we summarize the whole decoupling process for the distributed problem in Fig.\,\ref{decoupling process}. Specifically, to transform the coordinated form ${\Bar{R}_{c,k}^{\textnormal{bs}}\left(\Bar{\mathbf{f}},\Bar{\mathbf{v}}\right)}$ and ${\Bar{R}_{p,k}^{\textnormal{bs}}\left(\Bar{\mathbf{f}},\Bar{\mathbf{v}}\right)}$ into the distributed form $\Bar{\Gamma}_{c,k}^{\textnormal{bs}}\left(\bar{\mathbf{f}}\right)$ and $\Bar{\Gamma}_{p,k}^{\textnormal{bs}}\left(\bar{\mathbf{v}}\right)$, there are three major steps.
First, using the ergodic spectral efficiency and Jensen's inequality, we averaged the problematic IUI term $\Bar{\mathbf{v}}^{\sf H}\mathbf{U}_{p,k}^{\textnormal{bs}}\Bar{\mathbf{v}}$ of ${\Bar{R}_{p,k}^{\textnormal{bs}}\left(\Bar{\mathbf{f}},\Bar{\mathbf{v}}\right)}$ and the IUI term $\Bar{\mathbf{v}}^{\sf H}\mathbf{U}_{c,k}^{\textnormal{bs}}\Bar{\mathbf{v}}$ of ${\Bar{R}_{c,k}^{\textnormal{bs}}\left(\Bar{\mathbf{f}},\Bar{\mathbf{v}}\right)}$, and captured them as constants, respectively. 
Second, for the decoupling process, it is necessary to measure $\mathbb{E} \big[ \Bar{\mathbf{v}}^{\sf H}\mathbf{U}_{p,k}^{\textnormal{bs}}\Bar{\mathbf{v}} \big]$ and $\mathbb{E} \big[ \Bar{\mathbf{v}}^{\sf H}\mathbf{U}_{c,k}^{\textnormal{bs}}\Bar{\mathbf{v}} \big]$ and report them to the LEO satellite. This is done by calculating $\mathbb{E} \big[ \Bar{\mathbf{v}}^{\sf H}\mathbf{U}_{p,k}^{\textnormal{bs}}\Bar{\mathbf{v}} \big]$ and $\mathbb{E} \big[ \Bar{\mathbf{v}}^{\sf H}\mathbf{U}_{c,k}^{\textnormal{bs}}\Bar{\mathbf{v}} \big]$ according to the interference report mechanisms introduced in Remark 5 and reporting them to the LEO satellite.
Third, we finally isolate the ICI term into leakage interference, separating the functions related to $\Bar{\mathbf{f}}$ and $\Bar{\mathbf{v}}$, thereby transforming them into a distributed form.
}}

\begin{remark} \normalfont \label{remark:report}
(Decoupling intuition and report mechanism)
A key idea of the spectral efficiency decoupling is converting the IUI term to a constant through averaging. 
By averaging the IUI term, however, it is evident that the spectral efficiency would degrade because it is infeasible to reflect the instantaneous IUI in designing ${\bf{ \bar {f}}}$. For example, in \eqref{re_R_pri_TU_approx}, instead of considering the instantaneous IUI term $\Bar{\mathbf{v}}^{\sf H}\mathbf{U}_{p,k}^{\textnormal{bs}}\Bar{\mathbf{v}}$, 
we take into account the averaged IUI term 
$\mathbb{E} \big[ \Bar{\mathbf{v}}^{\sf H}\mathbf{U}_{p,k}^{\textnormal{bs}}\Bar{\mathbf{v}} \big]$ (defined as $\epsilon_k$).  
Hence, as increasing a disparity between $\Bar{\mathbf{v}}^{\sf H}\mathbf{U}_{p,k}^{\textnormal{bs}}\Bar{\mathbf{v}}$ and $\mathbb{E} \big[ \Bar{\mathbf{v}}^{\sf H}\mathbf{U}_{p,k}^{\textnormal{bs}}\Bar{\mathbf{v}} \big]$, the performance degradation is unavoidable. We understand that this is a price to pay for distributed precoding.  

We can come up with a way to compensate this performance degradation. 
For instance, if we are able to deliver the instantaneous IUI to the LEO satellite, it is possible to use $\Bar{\mathbf{v}}^{\sf H}\mathbf{U}_{p,k}^{\textnormal{bs}}\Bar{\mathbf{v}}$ as $\epsilon_k$ in \eqref{er_R_p_u}, by which more exact IUI is incorporated. 
Obviously, however, reporting the instantaneous IUI to the LEO satellite induces more overheads. 
On the contrary, if we want to completely eliminate the overheads associated with the IUI, we can set $0 \rightarrow \epsilon_k $ in \eqref{er_R_p_u}, though this would lead to more severe performance degradation. 
This intuition naturally leads to the following three different interference report mechanisms. We note that all of the following report mechanisms use the same distributed precoding method. The only difference is whether using the instantaneous IUI, the averaged IUI, or not using any IUI information.



\lowercase\expandafter{\romannumeral1}) {\textbf{Average report mechanism}}: 
In this case, 
we calculate the averaged IUI terms $\mathbb{E} \big[ \Bar{\mathbf{v}}^{\sf H}\mathbf{U}_{p,k}^{\textnormal{bs}}\Bar{\mathbf{v}} \big]$ and $\mathbb{E} \big[ \Bar{\mathbf{v}}^{\sf H}\mathbf{U}_{c,k}^{\textnormal{bs}}\Bar{\mathbf{v}} \big] $ in a Monte-Carlo fashion then report these to the LEO satellite for one time. In Section \uppercase\expandafter{\romannumeral5}, we will explain how to obtain $\mathbb{E} \big[ \Bar{\mathbf{v}}^{\sf H}\mathbf{U}_{p,k}^{\textnormal{bs}}\Bar{\mathbf{v}} \big]$ and $\mathbb{E} \big[ \Bar{\mathbf{v}}^{\sf H}\mathbf{U}_{c,k}^{\textnormal{bs}}\Bar{\mathbf{v}} \big] $ in details. 
As a consequence, we let $\mathbb{E} \big[ \Bar{\mathbf{v}}^{\sf H}\mathbf{U}_{p,k}^{\textnormal{bs}}\Bar{\mathbf{v}} \big] \rightarrow \epsilon_k$ and $\mathbb{E} \big[ \Bar{\mathbf{v}}^{\sf H}\mathbf{U}_{c,k}^{\textnormal{bs}}\Bar{\mathbf{v}} \big] \rightarrow \omega_k$ in \eqref{er_R_p_u} and \eqref{gamma_R_c_k}, respectively. 
Note that this method corresponds to a case that we mainly explain in Section \uppercase\expandafter{\romannumeral3}. 
This mechanism strikes a balance between the spectral efficiency performance and the amount of overheads. 

\lowercase\expandafter{\romannumeral2}) {\textbf{Instantaneous report mechanism}}:
In this case, we obtain the instantaneous IUI terms $ {\bf{\bar v}}^{\sf H} \mathbf{U}_{p,k}^{\textnormal{bs}} {\bf{\bar v}}$ and $\Bar{\mathbf{v}}^{\sf H}\mathbf{U}_{c,k}^{\textnormal{bs}}\Bar{\mathbf{v}}$ and report them 
to the LEO satellite. 
Since $ {\bf{\bar v}}^{\sf H} \mathbf{U}_{p,k}^{\textnormal{bs}} {\bf{\bar v}}$ and $\Bar{\mathbf{v}}^{\sf H}\mathbf{U}_{c,k}^{\textnormal{bs}}\Bar{\mathbf{v}}$ are changed depending on the instantaneous channel realization, the TUs are required to deliver them per every channel block, which requires more overheads. 
Consequently, we let $ \Bar{\mathbf{v}}^{\sf H}\mathbf{U}_{p,k}^{\textnormal{bs}}\Bar{\mathbf{v}}  \rightarrow \epsilon_k$ and $ \Bar{\mathbf{v}}^{\sf H}\mathbf{U}_{c,k}^{\textnormal{bs}}\Bar{\mathbf{v}}  \rightarrow \omega_k$ in \eqref{er_R_p_u} and \eqref{gamma_R_c_k}, respectively. This mechanism achieves the best spectral efficiency, while requiring large amount of overheads. 

\lowercase\expandafter{\romannumeral3}) {\textbf{Zero report mechanism}}: 
In this case, we never use the report associated with the IUI. 
Accordingly, we have $0 \rightarrow \epsilon_k $ and $0 \rightarrow \omega_k$ in \eqref{er_R_p_u} and \eqref{gamma_R_c_k}, respectively. While this mechanism has an advantage of not consuming overheads, it comes with a drawback of not reflecting the IUI, leading to the worst spectral efficiency performance. 

In Section \uppercase\expandafter{\romannumeral5}, we will numerically compare the spectral efficiencies for the three report mechanisms.
\end{remark}

\section{Precoder Optimization with Generalized Power Iteration}
In this section, we propose an algorithm to solve $\mathscr{P}_2$.

\subsection{LSE Approximation}
One challenge in solving $\mathscr{P}_2$ is the non-smoothness of the minimum function $ \min_{k \in \CMcal{K}_t^{{\text{int}}}, u \in \CMcal{K}_s} \big\{ \Bar{\Gamma}_{c,k}^{\textnormal{bs}} \left(\bar{\mathbf{f}}\right), \Bar{R}_{c,u}^{\textnormal{sat}}\left(\Bar{\mathbf{f}}\right)\big\}$.  
To resolve this, we first approximate the non-smooth minimum function in \eqref{re_opt_prob} as a smooth function using the LSE technique \cite{Shen:tpami:10}. Applying the LSE technique, the minimum function in \eqref{re_opt_prob} is approximated as follows
{\color{black}{
\begin{align}
    &\min_{k \in \CMcal{K}_t^{{\text{int}}}, u \in \CMcal{K}_s} \left\{\Bar{\Gamma}_{c,k}^{\textnormal{bs}}\left(\bar{\mathbf{f}}\right), \Bar{R}_{c,u}^{\textnormal{sat}}\left(\Bar{\mathbf{f}}\right)\right\}   \nonumber\\
    &\simeq - \mu \log\left[ \frac{1}{K^{{\text{sat}}}} \left\{\sum_{j \in \CMcal{K}_t^{{\text{int}}}}\exp\left( \frac{\Bar{\Gamma}_{c,j}^{\textnormal{bs}}\left(\bar{\mathbf{f}}\right)}{-\mu} \right) + \sum_{i=1}^{K_s} \exp\left( \frac{\Bar{R}_{c,i}^{\textnormal{sat}}\left(\bar{\mathbf{f}}\right)}{-\mu} \right)\right\} \right] \nonumber\\
    &=\Lambda\left(\bar{\mathbf{f}}\right),\label{LSE}
\end{align}}}
{\color{black}{where $\mu$ determines the accuracy of the LSE technique. As we set small $\mu$, \eqref{LSE} becomes tight. If convergence is not achieved within a predefined number of iterations as SNR changes, the value of $\mu$ is increased.}}
This leads to the following problem: 
\begin{align}
    \mathscr{P}_3: \; \underset{\bar{\mathbf{f}},\bar{\mathbf{v}}}{\textnormal{maximize}} \;\;
    {\color{black}{
    \left[ \Lambda\left(\bar{\mathbf{f}}\right) + \sum_{i=1}^{K_s} \Bar{\Gamma}_{p,i}^{\textnormal{sat}}\left(\bar{\mathbf{f}}\right) + \sum_{j=1}^{K_t} \Bar{\Gamma}_{p,j}^{\textnormal{bs}}\left(\bar{\mathbf{v}}\right)
    \right]
    }}
    .\label{opt_prob_leak}
\end{align}

\subsection{First-Order Optimality Condition}
To approach the solution for $\mathscr{P}_3$, we represent the Lagrangian function of \eqref{opt_prob_leak} as 
{\color{black}{
\begin{align}
        f\left(\bar{\mathbf{f}},\bar{\mathbf{v}}\right) &= \Bigg[ \underbrace{\Lambda\left(\bar{\mathbf{f}}\right) + \sum_{i=1}^{K_s} \Bar{\Gamma}_{p,i}^{\textnormal{sat}}\left(\bar{\mathbf{f}}\right)}_{f_1\left(\bar{\mathbf{f}}\right)} +  \underbrace{\sum_{j=1}^{K_t} \Bar{\Gamma}_{p,j}^{\textnormal{bs}}\left(\bar{\mathbf{v}}\right)}_{f_2\left(\bar{\mathbf{v}}\right)} \Bigg].\label{lag_func}
\end{align}
}}
We divide $f\left(\bar{\mathbf{f}},\bar{\mathbf{v}}\right)$ into two parts, wherein 
the first part $f_1\left(\bar{\mathbf{f}}\right)$ is related to $\bar{\mathbf{f}}$ and the second part $f_2\left(\bar{\mathbf{v}}\right)$ is related to $\bar{\mathbf{v}}$, respectively. We derive the first-order optimality conditions in the following lemma.  
\begin{lemma} \label{lem:1}
First, the first-order KKT condition of $f_1 (\bar {\bf{f}})$ is satisfied if 
\begin{align}
    &\mathbf{B}^{-1}\left(\bar{\mathbf{f}}\right) \mathbf{A}\left(\bar{\mathbf{f}}\right)\bar{\mathbf{f}} = \lambda^{\text{sat}}\left(\bar{\mathbf{f}}\right)\bar{\mathbf{f}},\label{KKT_sat}
\end{align}
{\color{black}{
where $\lambda^{\text{sat}}\left(\bar{\mathbf{f}}\right)$ is given by
\begin{align}
    \lambda^{\text{sat}}\left(\bar{\mathbf{f}}\right) &= \left[ e^{\frac{\Lambda\left(\bar{\mathbf{f}}\right)}{\log_2 e}} \cdot \prod_{i=1}^{K_s} \Bar{\Gamma}_{p,i}^{\textnormal{sat}}\left(\bar{\mathbf{f}}\right) \right] = \frac{\lambda_{1}^{\text{sat}} \left(\bar{\mathbf{f}}\right)}{\lambda_{2}^{\text{sat}}\left(\bar{\mathbf{f}}\right)},\label{lambda_sat}
\end{align}
and $\mathbf{A}\left(\bar{\mathbf{f}}\right)$ and $\mathbf{B}\left(\bar{\mathbf{f}}\right)$ are shown at the top of the next page.
}}

\begin{figure*}
{\color{black}{
    \begin{align}
        \mathbf{A}\left(\bar{\mathbf{f}}\right) 
        &= \lambda_{1}^{\text{sat}} \left(\bar{\mathbf{f}}\right) \cdot \sum\limits_{u=1}^{K_s} \left\{ \frac{\exp\left(-\frac{1}{\mu}\log_2\left( \frac{\bar{\mathbf{f}}^{\sf H}\mathbf{X}_{c,u}^{\text{sat}}\bar{\mathbf{f}}}{\bar{\mathbf{f}}^{\sf H}\mathbf{U}_{c,u}^{\text{sat}}\bar{\mathbf{f}}}\right)\right) \frac{\mathbf{X}_{c,u}^{\text{sat}}}{\bar{\mathbf{f}}^{\sf H}\mathbf{X}_{c,u}^{\text{sat}}\bar{\mathbf{f}}} }{\sum_{i=1}^{K_s}\exp\left(-\frac{1}{\mu}\log_2\left( \frac{\bar{\mathbf{f}}^{\sf H}\mathbf{X}_{c,i}^{\text{sat}}\bar{\mathbf{f}}}{\Bar{\mathbf{f}}^{\sf H}\mathbf{U}_{c,i}^{\text{sat}}\Bar{\mathbf{f}}}\right)\right)} + \frac{ \mathbf{S}_{p,u}^{\textnormal{sat}} + \mathbf{U}_{p,u}^{\textnormal{sat}} }{\Bar{\mathbf{f}}^{\sf H}\left( \mathbf{S}_{p,u}^{\textnormal{sat}} + \mathbf{U}_{p,u}^{\textnormal{sat}} \right) \Bar{\mathbf{f}} \cdot L\left(\Bar{\mathbf{f}}\right) } + \frac{ L\left(\Bar{\mathbf{f}}\right) }{ \nabla_{\Bar{\mathbf{f}}} L\left(\Bar{\mathbf{f}}\right) \Bar{\mathbf{f}} } \right\} \nonumber\\
        &+ \lambda_{1}^{\text{sat}} \left(\bar{\mathbf{f}}\right) \cdot \sum\limits_{k \in \CMcal{K}_t^{{\text{int}}}}\left\{\frac{\exp\left(-\frac{1}{\mu}\log_2 \left( \frac{ \bar{\mathbf{f}}^{\sf H} \mathbf{X}_{c,k}^{\text{bs}} \bar{\mathbf{f}}}{ \bar{\mathbf{f}}^{\sf H} \mathbf{Y}_{c,k}^{\text{bs}} \bar{\mathbf{f}}} \right) \right)  \frac{ \mathbf{X}_{c,k}^{\text{bs}} }{ \bar{\mathbf{f}}^{\sf H} \mathbf{X}_{c,k}^{\text{bs}} \bar{\mathbf{f}}} }{\sum_{j \in \CMcal{K}_t^{{\text{int}}}} \exp\left(-\frac{1}{\mu}\log_2\left( \frac{ \bar{\mathbf{f}}^{\sf H} \mathbf{X}_{c,j}^{\text{bs}} \bar{\mathbf{f}}}{ \bar{\mathbf{f}}^{\sf H} \mathbf{Y}_{c,j}^{\text{bs}} \bar{\mathbf{f}}} \right) \right)} \right\} ,\label{A}\\
        \mathbf{B}\left(\bar{\mathbf{f}}\right) 
        &= \lambda_{2}^{\text{sat}}\left(\bar{\mathbf{f}}\right) \cdot \left[ \sum\limits_{u=1}^{K_s} \left\{ \frac{\exp\left(-\frac{1}{\mu}\log_2\left( \frac{\bar{\mathbf{f}}^{\sf H}\mathbf{X}_{c,u}^{\text{sat}}\bar{\mathbf{f}}}{\bar{\mathbf{f}}^{\sf H}\mathbf{U}_{c,u}^{\text{sat}}\bar{\mathbf{f}}}\right)\right)  \frac{\mathbf{U}_{c,u}^{\text{sat}}}{\bar{\mathbf{f}}^{\sf H}\mathbf{U}_{c,u}^{\text{sat}}\bar{\mathbf{f}}} }{\sum_{i=1}^{K_s}\exp\left(-\frac{1}{\mu}\log_2\left( \frac{\bar{\mathbf{f}}^{\sf H}\mathbf{X}_{c,i}^{\text{sat}}\bar{\mathbf{f}}}{\Bar{\mathbf{f}}^{\sf H}\mathbf{U}_{c,i}^{\text{sat}}\Bar{\mathbf{f}}}\right)\right)} + \frac{ \mathbf{U}_{p,u}^{\textnormal{sat}} }{\Bar{\mathbf{f}}^{\sf H} \mathbf{U}_{p,u}^{\textnormal{sat}} \Bar{\mathbf{f}} \cdot L\left(\Bar{\mathbf{f}}\right) }
     \right\} + \sum\limits_{k \in \CMcal{K}_t^{{\text{int}}}} \left\{\frac{\exp\left(-\frac{1}{\mu}\log_2 \left( \frac{ \bar{\mathbf{f}}^{\sf H} \mathbf{X}_{c,k}^{\text{bs}} \bar{\mathbf{f}}}{ \bar{\mathbf{f}}^{\sf H} \mathbf{Y}_{c,k}^{\text{bs}} \bar{\mathbf{f}}} \right) \right) \frac{ \mathbf{Y}_{c,k}^{\text{bs}} }{ \bar{\mathbf{f}}^{\sf H} \mathbf{Y}_{c,k}^{\text{bs}} \bar{\mathbf{f}} } }{\sum_{j \in \CMcal{K}_t^{{\text{int}}}} \exp\left(-\frac{1}{\mu}\log_2\left( \frac{ \bar{\mathbf{f}}^{\sf H} \mathbf{X}_{c,j}^{\text{bs}} \bar{\mathbf{f}}}{ \bar{\mathbf{f}}^{\sf H} \mathbf{Y}_{c,j}^{\text{bs}} \bar{\mathbf{f}}} \right) \right)} \right\} \right]. \label{B}
    \end{align}}}
    \hrule
\end{figure*}

Next, the first-order KKT condition of $f_2(\bar {\bf{v}})$ is satisfied if 
\begin{align}
    \mathbf{D}^{-1}\left(\bar{\mathbf{v}}\right)\mathbf{C}\left(\bar{\mathbf{v}}\right)\bar{\mathbf{v}} = \lambda^{\text{bs}}\left(\bar{\mathbf{v}}\right) \bar{\mathbf{v}},\label{KKT_bs}
\end{align}
where $\lambda^{\text{bs}}\left(\bar{\mathbf{v}}\right)$, $\mathbf{C}\left(\Bar{\mathbf{v}}\right)$, and  $\mathbf{D}\left(\Bar{\mathbf{v}}\right)$ are given by
\begin{align}
    \lambda^{\text{bs}}\left(\bar{\mathbf{v}}\right) 
    = \prod_{j=1}^{K_t} \Bar{\Gamma}_{p,j}^{\textnormal{bs}}\left(\bar{\mathbf{v}}\right) 
    &= \frac{\lambda_{1}^{\text{bs}} \left(\bar{\mathbf{v}}\right)}{\lambda_{2}^{\text{bs}} \left(\bar{\mathbf{v}}\right)},\label{lambda_bs}
\end{align}
{\color{black}{
\begin{align}
    \mathbf{C}\left(\Bar{\mathbf{v}}\right) &= \lambda_{1}^{\text{bs}} \left(\bar{\mathbf{v}}\right) \cdot \sum_{j=1}^{K_t}\left(\frac{\mathbf{S}_{p,j}^{\textnormal{bs}}}{\Bar{\mathbf{v}}^{\sf H}\mathbf{S}_{p,j}^{\textnormal{bs}}\Bar{\mathbf{v}}}\right),\label{C}\\
    \mathbf{D}\left(\Bar{\mathbf{v}}\right) &= \lambda_{2}^{\text{bs}} \left(\bar{\mathbf{v}}\right) \cdot \sum_{j=1}^{K_t}\left(\frac{\mathbf{U}_{p,j}^{\textnormal{bs}}}{\Bar{\mathbf{v}}^{\sf H}\mathbf{U}_{p,j}^{\textnormal{bs}}\Bar{\mathbf{v}}}\right).\label{D}
\end{align}
\begin{myproof}
See Appendix A.
\end{myproof}}}
\end{lemma}

In Lemma \ref{lem:1}, the first important observation is that the optimality conditions are decoupled into two parts, each of which is solely related to $\bar {\bf{f}}$ and $\bar {\bf{v}}$, respectively. To be specific, differentiating the Lagrangian function $f(\bar {\bf{f}}, \bar {\bf{v}})$ in \eqref{lag_func} with regard to $\bar {\bf{f}}$, we have $\frac{\partial f(\bar {\bf{f}}, \bar {\bf{v}})}{ \partial \bar {\bf{f}}} = \frac{\partial f_1(\bar {\bf{f}})}{ \partial \bar {\bf{f}}} = g_1 (\bar {\bf{f}})$, where $g_1 (\bar {\bf{f}})$ is independent to $\bar {\bf{v}}$. Similar to this, we have  $\frac{\partial f(\bar {\bf{f}}, \bar {\bf{v}})}{ \partial \bar {\bf{v}}} = \frac{\partial f_2(\bar {\bf{v}})}{ \partial \bar {\bf{v}}} = g_2 (\bar {\bf{v}})$, where $g_2 (\bar {\bf{v}})$ is independent to $\bar {\bf{f}}$. We note that this is done by our decoupling process described in Section \uppercase\expandafter{\romannumeral3}. Thanks to this decoupling, the functions related to $\bar {\bf{f}}$ and $\bar {\bf{v}}$ are independently computed in a distributed way.

Next we explain how to achieve the local optimal point for each of $\bar {\bf{f}}$ and $\bar {\bf{v}}$. 
Note that the first-order KKT condition \eqref{KKT_sat} (regarding $\bar {\bf{f}}$) and \eqref{KKT_bs} (regarding $\bar {\bf{v}}$) are cast as a form of NEPv \cite{park:twc:23}. 
Specifically, in \eqref{KKT_sat}, $\Bar{\mathbf{f}}$ behaves as an eigenvector of the eigenvector-dependent matrix $\mathbf{B}^{-1}\left(\bar{\mathbf{f}}\right)\mathbf{A}\left(\bar{\mathbf{f}}\right)$, and in \eqref{KKT_bs}, $\Bar{\mathbf{v}}$ acts as an eigenvector of the eigenvector-dependent matrix $\mathbf{D}^{-1}\left(\bar{\mathbf{v}}\right)\mathbf{C}\left(\bar{\mathbf{v}}\right)$. 
In this relation, the eigenvalue $\lambda^{\text{sat}}\left(\bar{\mathbf{f}}\right)$ given by \eqref{lambda_sat} corresponds to the $\bar{\mathbf{f}}$-related term of the Lagrangian function $f_1\left(\bar{\mathbf{f}}\right)$, while the eigenvalue $\lambda^{\text{bs}}\left(\bar{\mathbf{v}}\right)$ given by \eqref{lambda_bs} corresponds to the $\bar{\mathbf{v}}$-related term of the Lagrangian function $f_2\left(\bar{\mathbf{v}}\right)$.
As a result, if we find the principal eigenvectors of \eqref{KKT_sat} and \eqref{KKT_bs}, then we reach the maximum stationary point. In doing so, we maximize the objective function \eqref{opt_prob_leak}. 
Note that this NEPv problem is different from the classical eigenvalue problem in that the matrix is dependent on its eigenvector itself. In what follows, we propose an algorithm to find the principal eigenvectors of \eqref{KKT_sat} and \eqref{KKT_bs}.

\subsection{STIN-GPI Algorithm}
We propose a STIN-GPI algorithm based on \cite{park:twc:23}. 
The process of the proposed algorithm is iteratively updating $\bar{\mathbf{f}}$ and $\bar{\mathbf{v}}$ by using the power iteration, while the matrix is updated with the previously obtained $\bar{\mathbf{f}}$ and $\bar{\mathbf{v}}$.
In our method, STIN-GPI consists of two stages, where the first stage is performed in the LEO satellite for designing $\bar {\bf{f}}$ and the second stage is performed in the terrestrial BS for designing $\bar {\bf{v}}$. We mention that these two stages are totally departed; so that no outcome of one stage goes into the other stage as the input.
{\color{black}{For the precoding vectors $\bar{\mathbf{f}}^{\left(0\right)}$ and $\bar{\mathbf{v}}^{\left(0\right)}$, the precoding vector $\mathbf{f}_c$ combined with the common stream is initialized as a random vector, while the initial values of the precoding vectors $\left\{ \mathbf{f}_{p,1}, \cdots, \mathbf{f}_{p,K_s} \right\}$ and $\left\{ \mathbf{v}_{1}, \cdots, \mathbf{v}_{K_t} \right\}$ combined with the private stream are set using the maximum ratio transmission (MRT) method \cite{park:twc:23, kim:wcl:23}.}}
In the first stage, with the given $\bar{\mathbf{f}}^{(t-1)}$ obtained in the $\left(t-1\right)$-th iteration, the matrices $\mathbf{A}(\bar{\mathbf{f}}^{\left(t-1\right)})$ and $\mathbf{B}(\bar{\mathbf{f}}^{\left(t-1\right)})$ are calculated utilizing \eqref{A} and \eqref{B}. Then, we update $\bar {\bf{f}}^{(t)}$ in the $(t)$-th iteration by $\bar{\mathbf{f}}^{(t)} = \frac{\mathbf{B}^{-1}\left(\bar{\mathbf{f}}^{(t-1)}\right)\mathbf{A}\left(\bar{\mathbf{f}}^{(t-1)}\right)\bar{\mathbf{f}}^{(t-1)}}{\lVert\mathbf{B}^{-1}\left(\bar{\mathbf{f}}^{(t-1)}\right)\mathbf{A}\left(\bar{\mathbf{f}}^{(t-1)}\right)\bar{\mathbf{f}}^{(t-1)}\rVert}$.
This process repeats until convergence under a predetermined tolerance value $\zeta$, i.e., $\lVert\bar{\mathbf{f}}^{\left(t\right)}-\bar{\mathbf{f}}^{\left(t-1\right)}\rVert<\zeta$.
The second stage runs with the same way. The matrices $\mathbf{C}(\bar{\mathbf{v}}^{\left(t-1\right)})$ and $\mathbf{D}(\bar{\mathbf{v}}^{\left(t-1\right)})$ are calculated using $\bar{\mathbf{v}}^{(t-1)}$. Then it is updated by $\bar{\mathbf{v}}^{(t)} = \frac{\mathbf{D}^{-1}\left(\bar{\mathbf{v}}^{(t-1)}\right)\mathbf{C}\left(\bar{\mathbf{v}}^{(t-1)}\right)\bar{\mathbf{v}}^{(t-1)}}{\lVert\mathbf{D}^{-1}\left(\bar{\mathbf{v}}^{(t-1)}\right)\mathbf{C}\left(\bar{\mathbf{v}}^{(t-1)}\right)\bar{\mathbf{v}}^{(t-1)}\rVert}$.
This process repeats until $\lVert\bar{\mathbf{v}}^{\left(t\right)}-\bar{\mathbf{v}}^{\left(t-1\right)}\rVert<\zeta$. 
We describe the whole process of our proposed method in Algorithm 1.

We note that $\mu$ sets a trade-off between the performance and the convergence. That is to say, large $\mu$ is favorable as it better approximates the minimum function, but too large $\mu$ may lead to divergence due to a lack of smoothness. Accordingly, it is important to use a proper value of $\mu$. 
{\textcolor{black}{In the next section, we outline the process for finding a proper value of $\mu$ during the iteration procedure.
}}


\begin{algorithm}[t]
\caption{STIN-GPI algorithm}\label{alg:1}
\KwInput{$\bar{\mathbf{f}}^{\left(0\right)}=\left(\textnormal{MRT}\right)$, $\bar{\mathbf{v}}^{\left(0\right)}=\left(\textnormal{MRT}\right)$, $t=1$}
\Repeat{ $t = t^{\textnormal{max}}$ }{
    \textbf{Stage 1.}\,\,\textit{LEO Satellite Beamforming Design}\\
    \Repeat{$\lVert\bar{\mathbf{f}}^{\left(t\right)}-\bar{\mathbf{f}}^{\left(t-1\right)}\rVert<\zeta$}
    {\textnormal{Calculate} $\mathbf{A}(\bar{\mathbf{f}}^{\left(t-1\right)}),\mathbf{B}(\bar{\mathbf{f}}^{\left(t-1\right)})$\\
    \textnormal{Obtain} $\bar{\mathbf{f}}^{\left(t\right)} \leftarrow \frac{\mathbf{B}^{-1} (\bar{\mathbf{f}}^{(t-1)} )\mathbf{A} (\bar{\mathbf{f}}^{(t-1)} )\bar{\mathbf{f}}^{(t-1)}}{\lVert\mathbf{B}^{-1}(\bar{\mathbf{f}}^{(t-1)})\mathbf{A}(\bar{\mathbf{f}}^{(t-1)})\bar{\mathbf{f}}^{(t-1)}\rVert}$}    
    \textbf{Stage 2.}\,\,\textit{Terrestrial BS Beamforming Design}\\
    \Repeat{$\lVert\bar{\mathbf{v}}^{\left(t\right)}-\bar{\mathbf{v}}^{\left(t-1\right)}\rVert<\zeta$}
    {\textnormal{Calculate} $\mathbf{C}(\bar{\mathbf{v}}^{\left(t-1\right)}),\mathbf{D}(\bar{\mathbf{v}}^{\left(t-1\right)})$\\
    \textnormal{Obtain} $\bar{\mathbf{v}}^{\left(t\right)} \leftarrow \frac{\mathbf{D}^{-1}(\bar{\mathbf{v}}^{(t-1)})\mathbf{C}(\bar{\mathbf{v}}^{(t-1)})\bar{\mathbf{v}}^{(t-1)}}{\lVert\mathbf{D}^{-1}(\bar{\mathbf{v}}^{(t-1)})\mathbf{C}(\bar{\mathbf{v}}^{(t-1)})\bar{\mathbf{v}}^{(t-1)}\rVert}$}
    $t\leftarrow t+1$}
\KwOutput{$\bar{\mathbf{f}}^{\left(t^{\textnormal{max}}\right)}, \bar{\mathbf{v}}^{\left(t^{\textnormal{max}}\right)}$}
\end{algorithm}


\section{Numerical Results} \label{sec:num}
\begin{table}[t]
\centering
\caption{{\color{black}{Simulation Parameters}}}
\label{parameters}
\begin{tabular}{c|c}
\noalign{\smallskip}\noalign{\smallskip}\hline\hline
Parameters & Value\\
\hline
Topology of user & Uniformly distributed per cell \\
Carrier frequency & $f_c$ = 20 GHz \\
Bandwidth & $B_w$ = 800 MHz \\
Orbit altitude & $d_0^{\text{sat}}$ = 1000 km (LEO) \\
Terrestrial BS height & 30 m \\
Radius of LEO satellite coverage & 500 km \\
Radius of terrestrial BS coverage & 50 km \\
The number of LEO satellite antennas & $M_1=5, M_2=5, M=25$ \\
The number of terrestrial BS antennas & $N_1=3, N_2=3, N=9$ \\
The number of users & $K_s=10, K_t=3, K_t^{\text{int}}=1$ \\
Antenna gain & $G_{\textnormal{sat}} = 6$ dBi, $G_{u} = 0$ dBi \\
UPA inter-element spacing & $d_1^{\text{sat}} = d_2^{\text{sat}} = \lambda, d_1^{\text{bs}} = d_2^{\text{bs}} = \frac{\lambda}{2}$\\
Number of NLoS paths & $L_t=10$ \\
Pilot length and pilot transmission power & $\tau p^{\text{pi}}=2$ \\
{\color{black}{LoS/NLoS ratio}} & {\color{black}{$\kappa_s=10$}} \\
Path loss exponent (Terrestrial channel) & $\rho=4$ \\
Tolerance value for convergence & $\zeta=0.01$ \\
\hline
\hline
\end{tabular}
\end{table}
In this section, we evaluate the performance of the proposed method via numerical simulations. 
To this end, we consider the following STIN scenario. 
The STIN uses the Ka-band as operating bandwidth \cite{Li:tcom:22, You:jsac:22, 3gpp:38:811}. 
The radius of the LEO satellite coverage region is $500{\text{km}}$, while the radius of terrestrial BS coverage is $50{\text{km}}$ and the height of terrestrial BS is $30{\text{m}}$. SUs and TUs are uniformly distributed within their respective coverage areas.
{\color{black}{
The simulation setups are given in Table~\ref{parameters}} by adopting the simulation setups used in \cite{You:jsac:22, Li:tcom:22}. 
We use these parameters in the simulations unless mentioned otherwise. 
}
As baseline methods, we consider the followings: 
\begin{itemize}
    {\color{black}{\item \textbf{Coordinated Precoding with RS (Coord-RS)}:
    In this method, we find the optimal precoder to solve $\mathscr{P}_1$ in a coordinated fashion, i.e., by sharing CSIT, while incorporating the RS strategy. The basic setup of Coord-RS corresponds to \cite{yin:twc:23}, except that \cite{yin:twc:23} maximizes the minimum spectral efficiency while Coord-RS maximizes the sum spectral efficiency.}}
    {\item \textbf{CCCP} \cite{Li:jsac:20, Zhao:twc:23}}: This method approximates the RS precoding optimization problem using convex-concave procedure (CCCP) and solves it by using an off-the-shelf toolbox such as CVX. 
    We apply CCCP in the LEO satellite while ignoring the TUs.
    \item \textbf{SILNR Max} \cite{han:tcom:21}: This method adopts the SILNR instead of the exact SINR for distributed design. No RS is used. 
    \item \textbf{IUI-ICI Separation} \cite{Choi:twc:12}: This method decouples the IUI and the ICI based on the high SNR assumption. No RS is used. 
    \item \textbf{SLNR Max} \cite{sadek:twc:07}: The SLNR is adopted as an alternative to the exact SINR. No RS is used. 
    \item \textbf{Local ZF} \cite{Interdonato:twc}: This method projects a precoding vector to null space of IUI and ICI by using the remaining spatial degrees-of-freedom. No RS is used.
    \item \textbf{Single-cell ZF} : This method is classical zero-forcing (ZF) that only suppresses the IUI. No RS is used. 
\end{itemize}

We consider three different versions of the proposed STIN-GPI methods depending on the report mechanisms: STIN-GPI-Ins. (Instantaneous report mechanism), STIN-GPI-Avg. (Average report mechanism), and STIN-GPI-Zero. (Zero report mechanism), where the main differences are explained in Remark \ref{remark:report}. 
In STIN-GPI-Avg., to get $\mathbb{E} \big[ \Bar{\mathbf{v}}^{\sf H}\mathbf{U}_{p,k}^{\textnormal{bs}}\Bar{\mathbf{v}} \big]$ and $\mathbb{E} \big[ \Bar{\mathbf{v}}^{\sf H}\mathbf{U}_{c,k}^{\textnormal{bs}}\Bar{\mathbf{v}} \big]$, we form $1000$ samples of the channel vector for TU $k$, $k \in \CMcal{K}_t$ in a Monte-Carlo fashion and construct $\mathbf{U}_{p,k}^{\textnormal{bs}}$ and $\mathbf{U}_{c,k}^{\textnormal{bs}}$. 
Then, we design $\Bar{\mathbf{v}}$ according to Stage $2$ in Algorithm $1$ and obtain $\mathbb{E} \big[ \Bar{\mathbf{v}}^{\sf H}\mathbf{U}_{p,k}^{\textnormal{bs}}\Bar{\mathbf{v}} \big]$ and $\mathbb{E} \big[ \Bar{\mathbf{v}}^{\sf H}\mathbf{U}_{c,k}^{\textnormal{bs}}\Bar{\mathbf{v}} \big]$ by numerically averaging $ \Bar{\mathbf{v}}^{\sf H}\mathbf{U}_{p,k}^{\textnormal{bs}}\Bar{\mathbf{v}}$ and $ \Bar{\mathbf{v}}^{\sf H}\mathbf{U}_{c,k}^{\textnormal{bs}}\Bar{\mathbf{v}}$. Finally, we set them as $\mathbb{E} \big[ \Bar{\mathbf{v}}^{\sf H}\mathbf{U}_{p,k}^{\textnormal{bs}}\Bar{\mathbf{v}} \big] \rightarrow \epsilon_k$ and $\mathbb{E} \big[ \Bar{\mathbf{v}}^{\sf H}\mathbf{U}_{c,k}^{\textnormal{bs}}\Bar{\mathbf{v}} \big] \rightarrow \omega_k$, respectively.

Additionally, as presented in Remark \ref{remark:general}, we also consider the $2$-layer RS strategy that composes two types of common streams. For distributed precoding applying to the $2$-layer RS strategy, we use STIN-GPI-Avg. 
{\color{black}{
In addition, notice that all the plotted spectral efficiencies are computed based on the true channel. 
Specifically, each user measures the post-processing SINR and informs this to the satellite or the terrestrial BS through uplink control channel. Thus, the imperfect knowledge on CSIT only degrades the achievable spectral efficiency, but this does not cause outage events. 
}}

\begin{figure}[t]
 \renewcommand{\figurename}{Fig.}
    \centering
    \includegraphics[width=0.45\textwidth ]{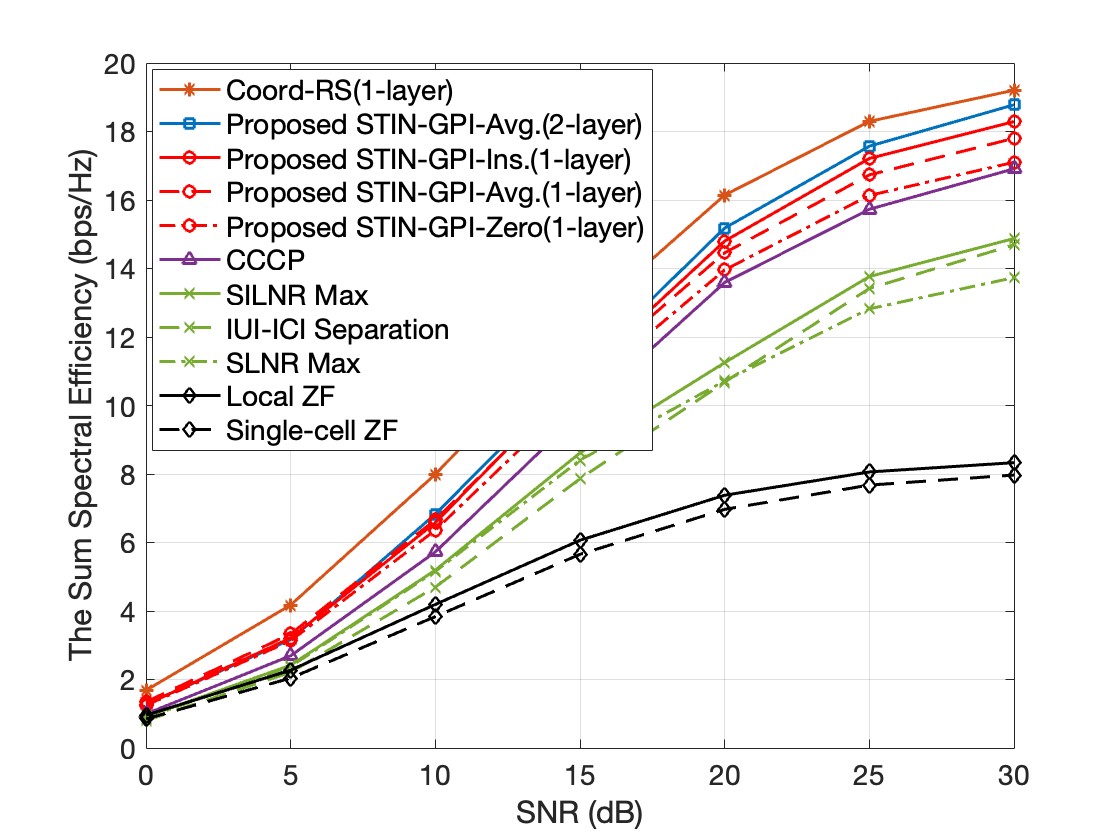}
    \caption{{\color{black}{Comparison of sum spectral efficiency between the proposed distributed STIN-GPI and baseline methods.}}}
    \label{sum_rate}
\end{figure}

{\textbf{The sum spectral efficiency}}: Fig.\,\ref{sum_rate} shows the ergodic sum spectral efficiency per SNR.
We first observe that, in the proposed methods, the spectral efficiency order is STIN-GPI-Ins. $>$ STIN-GPI-Avg. $>$ STIN-GPI-Zero. 
This is reasonable because STIN-GPI-Ins. exploits the exact IUI information, while STIN-GPI-Avg. only uses the averaged one and STIN-GPI-Zero does not use any information in designing precoders. 
However, STIN-GPI-Ins. requires instantaneous interference reporting from the TUs, incurring a substantial amount of overhead.
Unlike this, STIN-GPI-Avg. merely needs to have the constant $\epsilon_k$ and $\omega_k$ that does not vary over channels, thus the associated overheads are significantly smaller than those of STIN-GPI-Ins. Considering the trade-off between the performance gains and the associated overheads, STIN-GPI-Avg. is the most favorable option. 


Fig.\,\ref{sum_rate} shows that the STIN-GPI-Avg.(1-layer) offers around $20\%$ and $29\%$ gains at $\text{SNR}=30\text{dB}$ over the SILNR Max and the SLNR Max, respectively. 
These gains stem from two perspectives:
\lowercase\expandafter{\romannumeral1}) our method enables to perform SIC at the TUs by using the RS strategy. 
Given that the SILNR Max and the SLNR Max employ TIN-based decoding, we interpret the observed gains as evidence demonstrating the superiority of RS over TIN in a medium-strong interference regime, typically encountered in the considered STINs. 
\lowercase\expandafter{\romannumeral2}) Our decoupling technique uses meticulous treatment that establishes a lower bound on the spectral efficiency, aptly encompassing the interference's influence. 

{{\color{black}{
Comparing Coord-RS with STIN-GPI-Avg.(1-layer), there is a $7\%$ performance difference at $\text{SNR}=30\text{dB}$. This gap is attributed to the CSIT sharing between the LEO satellite and the terrestrial BS in Coord-RS. 
By using the instantaneous reporting mechanism (STIN-GPI-Ins.), 
the performance gap decreases to $5\%$. 
Further, employing the 2-layer RS strategy can further reduces this gap to $2\%$. 
This implies that aided by the instantaneous reporting and ICI management for TUs through the super common stream $s_{sc}^{\textnormal{sat}}$, our method achieves comparable performance to the coordinated approach even in the absence of CSIT sharing.}}
Compared to CCCP \cite{Li:jsac:20} that employs the RS strategy, we observe that the STIN-GPI-Avg.(1-layer) provides around $5\%$ gains at $\text{SNR}=30\text{dB}$. 
This improvement is attributed to the sophisticated decoupling method, which is not the case of CCCP \cite{Li:jsac:20}.
Moreover, the 2-layer RS strategy yields approximately $5.4\%$ gains over the 1-layer RS strategy at $\text{SNR}=30\text{dB}$. This is because the 2-layer RS strategy mitigates both ICI and IUI simultaneously through two types of common messages, thereby further enhancing the sum spectral efficiency at the cost of decoding complexity.

\begin{figure}[t]
\centering
    \subfigure[]{{\includegraphics[width=0.45\textwidth ]{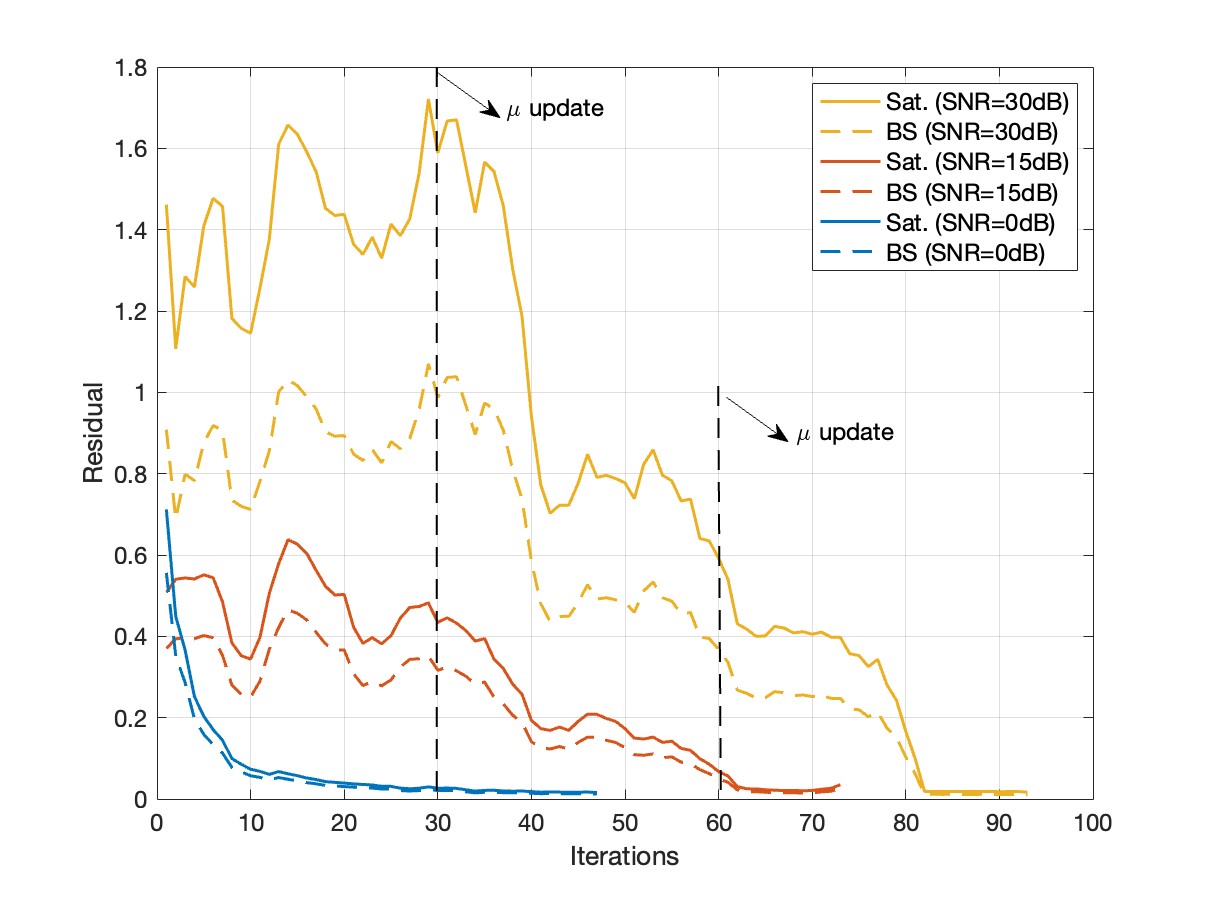}}
    \label{Convergence_v1}
    }
    \subfigure[]{{\includegraphics[width=0.45\textwidth ]{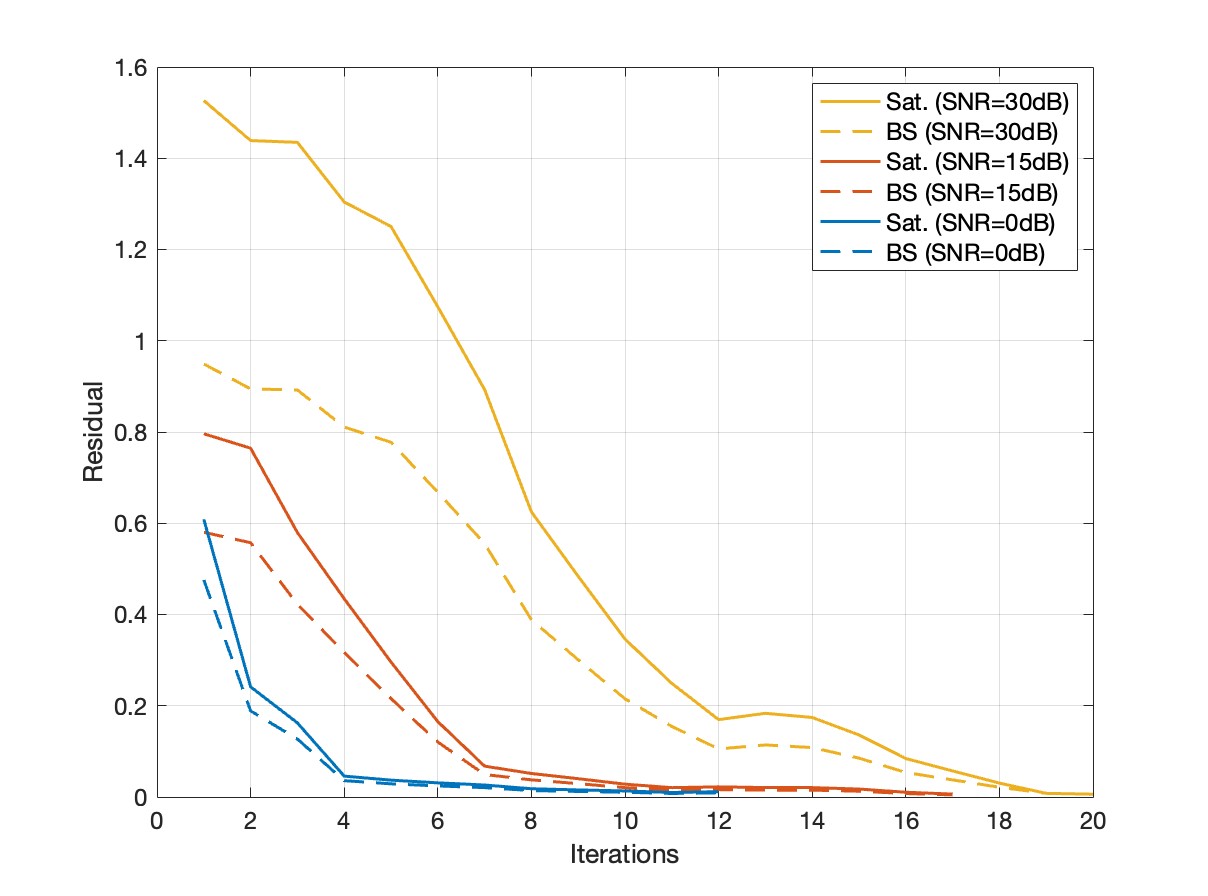}}
    \label{Convergence_v2}
    } 
\caption{{\color{black}{The residual convergence comparison as the iterations, $\text{SNR}=0,15,30\text{dB}$. (a) set the initial value $\mu=0.1$ (b) set by increasing the initial value $\mu=0.5,0.8,1$ at $\text{SNR}=0,15,30\text{dB}$, respectively.}}}
\label{Convergence} 
\end{figure}

{\color{black}{
{\textbf{$\mu$ update for fast convergence}}: As shown in Fig.\,\ref{Convergence}, we provide numerical results demonstrating the convergence behavior of the proposed algorithm.
For the detailed algorithm setup, we set the initial $\mu$ value to 0.1 and increase it by 0.2 unless the algorithm does not converge within 30 iterations. This process is designed to identify the smallest $\mu$ that guarantees convergence as explained above. 
In Fig.\,\ref{Convergence_v1}, we observe that the proposed method converges well with a small value of $\mu$ in the low SNR regime, such as $\text{SNR}=0\text{dB}$. 
In the high SNR regime, however, updating $\mu$ is required to ensure the convergence. 
For example, in the case of $\text{SNR}=15\text{dB}$, $\mu$ is updated once, and in the case of $\text{SNR}=30\text{dB}$, $\mu$ is adjusted twice until convergence. As a result, the associated complexity may increase because it takes the longer iteration times to converge.
To reduce the complexity further, we can adapt the initial value of $\mu$. As shown in Fig.\,\ref{Convergence_v2}, the initial $\mu$ for each SNR is set in a different way to achieve quicker convergence. 
To be specific, we set the initial $\mu=0.5,0.8,1$ at $\text{SNR}=0,15,30\text{dB}$, respectively. 
In general, higher SNR requires a large value of $\mu$. 
By doing this, the iterations of all the SNR cases are completed within 20 iterations as shown in Fig.\,\ref{Convergence_v2}, by which the associated complexity significantly decreases. 
}}

\begin{figure}[t]
\centering
    \subfigure[]{{\includegraphics[width=0.45\textwidth ]{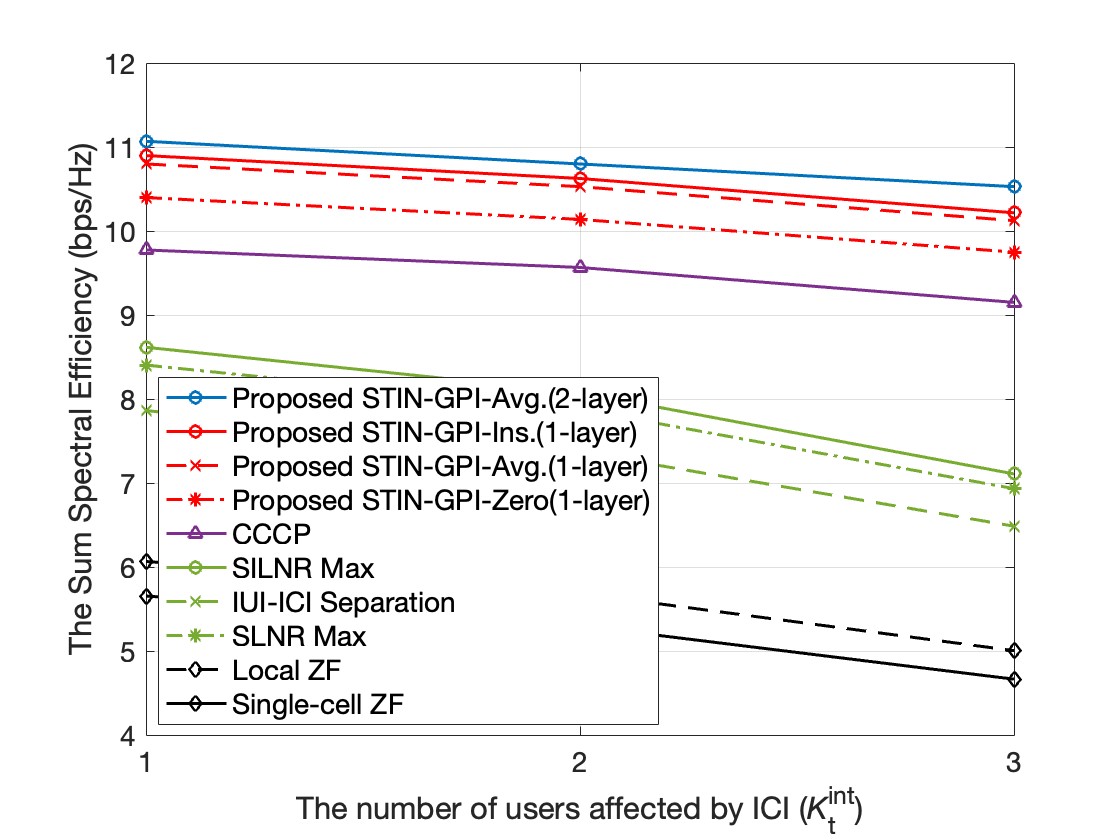}}
    \label{ici_user}
    } 
    \subfigure[]{{\includegraphics[width=0.45\textwidth ]{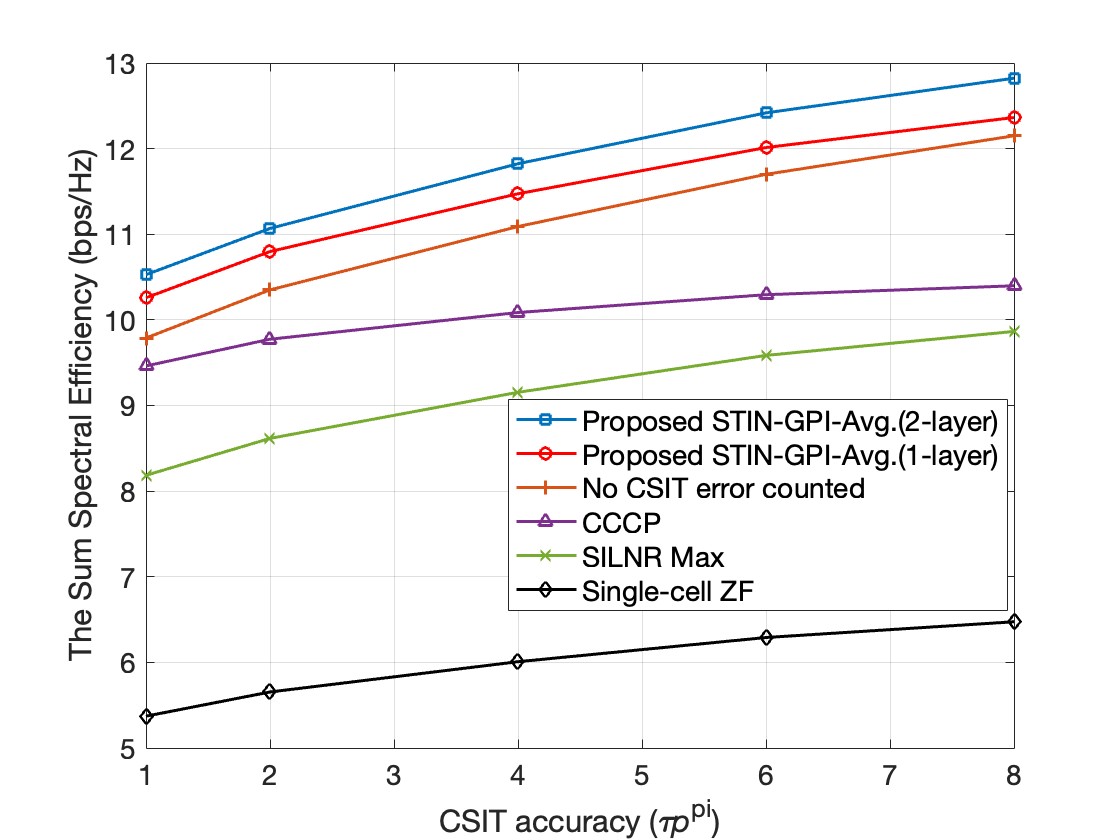}}
    \label{CSIT}
    }
\caption{Comparison of sum spectral efficiency among different strategies under the assumption of $\text{SNR}=15\text{dB}$ (a) per the number of TUs experiencing interference from the satellite $(K_t^{{\text{int}}})$, {\color{black}{(b) per the CSIT accuracy $(\tau p^{\textnormal{pi}})$.}}}
\label{comparison} 
\end{figure}

\begin{table}[t]
\centering
\caption{Comparison for the Computation Time (Seconds)}
\label{time_comparison}
\begin{tabular}{c|ccccc}
\toprule
\multicolumn{1}{c}{} & \multicolumn{3}{c}{$(K_s,K_t)$}\\
\cmidrule(rl){2-4}
\textbf{Method} & $(10,3)$ & $(15,5)$ & $(20,7)$\\
\midrule
STIN-GPI-Avg.(1-layer) & 5.95 (1\%) & 7.85 (1.3\%) & 12.05 (1.2\%)\\
STIN-GPI-Avg.(2-layer) & 11.21 (2\%) & 14.25 (2.4\%) & 20.57 (2.1\%)\\
CCCP & 416.59 & 586.13 & 971.60\\
\bottomrule
\end{tabular}
\end{table}

{\textbf{Computation time}}: 
In Table~\ref{time_comparison}, we compare the computation time between the proposed STIN-GPI and CCCP \cite{Li:jsac:20} as a rough measure of complexity. As observed in Table~\ref{time_comparison}, the proposed STIN-GPI-Avg. only consumes $1 \sim 2\%$ computation time compared to CCCP. 
This complexity reduction is attained by the proposed GPI-based optimization algorithm, which does not rely on an off-the-shelf optimization toolbox such as CVX.

\textbf{Per $K_t^{\textnormal{int}}$ and CSIT accuracy $\tau p^{\textnormal{pi}}$}: Fig.\,\ref{ici_user} compares the sum spectral efficiency per the number of TUs $K_t^{{\text{int}}}$ that experiences the interference from the LEO satellite.
We observe that, as $K_t^{{\text{int}}}$ increases, 
the proposed STIN-GPI-Ins. method decreases by $7\%$, while other methods (SILNR Max, IUI-ICI Separation, SLNR Max) decrease by $28\%, 33\%$, and $39\%$, respectively. 
The rationale behind this observation lies in capability of the RS strategy, that effectively mitigates the ICI, offering robustness against the increased ICI.

To further investigate the performance depending on the CSIT accuracy, in Fig.\,\ref{CSIT}, we compare the sum spectral efficiency depending on $\tau p^{\textnormal{pi}}$. 
As increasing $\tau p^{\textnormal{pi}}$, the CSIT estimation accuracy also increases. We observe STIN-GPI-Avg. outperforms CCCP in an entire region of $\tau p^{\textnormal{pi}}$. 
To be specific, as $\tau p^{\textnormal{pi}}$ increases from $1$ to $8$,
STIN-GPI-Avg.(1-layer) shows a $20\%$ improvement, while the performance of CCCP improves by $9\%$. This indicates that the relative performance gains of STIN-GPI-Avg. over CCCP increase as the CSIT becomes more accurate. 
{\color{black}{
Furthermore, to analyze the performance with and without reflecting CSIT estimation error, we compared the case assuming perfect channel estimation with our performance. If channel estimation is perfect, estimation error is not considered, and thus it follows $\mathbf{\Phi}_{k}^{\textnormal{sat}}= \mathbf{\Phi}_{k}^{\textnormal{bs}}=\mathbf{\Psi}_{u}^{\textnormal{sat}}=0$.
As shown in Fig.\,\ref{CSIT}, we observed that STIN-GPI-Avg.(1-layer) outperforms No CSIT error counted across the entire $\tau p^{\textnormal{pi}}$ range. Specifically, when $\tau p^{\textnormal{pi}}=1$, STIN-GPI-Avg.(1-layer) shows a $4.8\%$ performance improvement over No CSIT error counted method, and when $\tau p^{\textnormal{pi}}=8$, it shows a $1.7\%$ performance improvement.
}}
This arises from the fact that the CSIT estimation error is suitably reflected in the STIN-GPI design.


\section{Conclusion}
In this paper, we propose a novel distributed precoding approach using RS strategy. 
Our key idea is to decouple the sum spectral efficiency into two separated terms, each of which is a sole function of the satellite's precoder and the terrestrial BS's precoder, respectively. 
Based on the obtained distributed optimization problem, we approximate the non-smooth objective function by using the LSE technique, thereafter develop the STIN-GPI algorithm that finds the best local optimal point. 
The simulation results show that the proposed STIN-GPI method achieves considerable performance gains over the existing methods and it is suitable to be used for handling the interference in STINs. 
In future work, it is promising to extend this work by incorporating a multi-satellites and multi-BSs environment. 
Additionally, it is also interesting to develop a distributed transmission for security \cite{lee:tcom:22} in the STIN scenario.

\section*{Appendix A\\Proof of Lemma 1}
{\color{black}{
To find a stationary point of \eqref{lag_func}, we split the process into two steps. In the first step, we take the partial derivative of $f_1\left(\bar{\mathbf{f}}\right)$ with respect to $\bar{\mathbf{f}}$ and set it to zero. In the second step, we take the partial derivative of $f_2\left(\bar{\mathbf{v}}\right)$ with respect to $\bar{\mathbf{v}}$ and set it to zero as follows
\begin{align}
    &\frac{\partial f_1\left(\bar{\mathbf{f}}\right)}{\partial \bar{\mathbf{f}}} = 0,\label{deriv_f1}\\
    &\frac{\partial f_2\left(\bar{\mathbf{v}}\right)}{\partial \bar{\mathbf{v}}} = 0.\label{deriv_f2}
\end{align}
Specifically, by using 
\begin{align}
    \partial\left(\frac{\bar{\mathbf{f}}^{\sf H}\mathbf{A}\bar{\mathbf{f}}}{\bar{\mathbf{f}}^{\sf H}\mathbf{B}\bar{\mathbf{f}}}\right) \bigg/ \partial \bar{\mathbf{f}}^{\sf H}
     = \left(\frac{\bar{\mathbf{f}}^{\sf H}\mathbf{A}\bar{\mathbf{f}}}{\bar{\mathbf{f}}^{\sf H}\mathbf{B}\bar{\mathbf{f}}}\right) \left[\frac{\mathbf{A}\bar{\mathbf{f}}}{\bar{\mathbf{f}}^{\sf H}\mathbf{A}\bar{\mathbf{f}}} - \frac{\mathbf{B}\bar{\mathbf{f}}}{\bar{\mathbf{f}}^{\sf H}\mathbf{A}\bar{\mathbf{f}}}\right],
\end{align}
\begin{figure*}
    {\color{black}{
    \begin{align}
        \frac{\partial f_1\left(\bar{\mathbf{f}}\right)}{\partial \bar{\mathbf{f}}} 
        &= \frac{1}{\log 2} \sum\limits_{u=1}^{K_s} \left\{ \frac{ \left( \mathbf{S}_{p,u}^{\textnormal{sat}} + \mathbf{U}_{p,u}^{\textnormal{sat}} \right) \Bar{\mathbf{f}} }{\Bar{\mathbf{f}}^{\sf H}\left( \mathbf{S}_{p,u}^{\textnormal{sat}} + \mathbf{U}_{p,u}^{\textnormal{sat}} \right) \Bar{\mathbf{f}} \cdot L\left(\Bar{\mathbf{f}}\right) } - \frac{ \mathbf{U}_{p,u}^{\textnormal{sat}} \Bar{\mathbf{f}} }{\Bar{\mathbf{f}}^{\sf H} \mathbf{U}_{p,u}^{\textnormal{sat}} \Bar{\mathbf{f}} \cdot L\left(\Bar{\mathbf{f}}\right) } 
        + \frac{ L\left(\Bar{\mathbf{f}}\right) }{ \nabla_{\Bar{\mathbf{f}}} L\left(\Bar{\mathbf{f}}\right) }
        \right\}\nonumber\\
        &+\sum\limits_{k \in \CMcal{K}_t^{{\text{int}}}} \left\{\frac{\exp\left(-\frac{1}{\mu}\log_2 \left( \frac{ \bar{\mathbf{f}}^{\sf H} \mathbf{X}_{c,k}^{\text{bs}} \bar{\mathbf{f}}}{ \bar{\mathbf{f}}^{\sf H} \mathbf{Y}_{c,k}^{\text{bs}} \bar{\mathbf{f}}} \right) \right) \left( \frac{ \mathbf{X}_{c,k}^{\text{bs}} \bar{\mathbf{f}} }{  \bar{\mathbf{f}}^{\sf H} \mathbf{X}_{c,k}^{\text{bs}} \bar{\mathbf{f}}} - \frac{ \mathbf{Y}_{c,k}^{\text{bs}} \bar{\mathbf{f}}}{ \bar{\mathbf{f}}^{\sf H} \mathbf{Y}_{c,k}^{\text{bs}} \bar{\mathbf{f}}} \right) }{\log 2 \cdot \sum_{j \in \CMcal{K}_t^{{\text{int}}}} \exp\left(-\frac{1}{\mu}\log_2\left( \frac{ \bar{\mathbf{f}}^{\sf H} \mathbf{X}_{c,j}^{\text{bs}} \bar{\mathbf{f}}}{ \bar{\mathbf{f}}^{\sf H} \mathbf{Y}_{c,j}^{\text{bs}} \bar{\mathbf{f}}} \right) \right)} \right\} 
        + \sum\limits_{u=1}^{K_s} \left\{ \frac{\exp\left(-\frac{1}{\mu}\log_2\left( \frac{\bar{\mathbf{f}}^{\sf H}\mathbf{X}_{c,u}^{\text{sat}}\bar{\mathbf{f}}}{\bar{\mathbf{f}}^{\sf H}\mathbf{U}_{c,u}^{\text{sat}}\bar{\mathbf{f}}}\right)\right) \left(\frac{\mathbf{X}_{c,u}^{\text{sat}}\bar{\mathbf{f}}}{\bar{\mathbf{f}}^{\sf H}\mathbf{X}_{c,u}^{\text{sat}}\bar{\mathbf{f}}} - \frac{\mathbf{U}_{c,u}^{\text{sat}}\bar{\mathbf{f}}}{\bar{\mathbf{f}}^{\sf H}\mathbf{U}_{c,u}^{\text{sat}}\bar{\mathbf{f}}}\right)}{\log 2 \cdot \sum_{i=1}^{K_s}\exp\left(-\frac{1}{\mu}\log_2\left( \frac{\bar{\mathbf{f}}^{\sf H}\mathbf{X}_{c,i}^{\text{sat}}\bar{\mathbf{f}}}{\Bar{\mathbf{f}}^{\sf H}\mathbf{U}_{c,i}^{\text{sat}}\Bar{\mathbf{f}}}\right)\right)} \right\}.\label{L1_deriv}
    \end{align}
    \begin{align}
        &\sum\limits_{u=1}^{K_s} \left\{ \frac{\exp\left(-\frac{1}{\mu}\log_2\left( \frac{\bar{\mathbf{f}}^{\sf H}\mathbf{X}_{c,u}^{\text{sat}}\bar{\mathbf{f}}}{\bar{\mathbf{f}}^{\sf H}\mathbf{U}_{c,u}^{\text{sat}}\bar{\mathbf{f}}}\right)\right) \frac{\mathbf{X}_{c,u}^{\text{sat}}\bar{\mathbf{f}}}{\bar{\mathbf{f}}^{\sf H}\mathbf{X}_{c,u}^{\text{sat}}\bar{\mathbf{f}}} }{ \sum_{i=1}^{K_s}\exp\left(-\frac{1}{\mu}\log_2\left( \frac{\bar{\mathbf{f}}^{\sf H}\mathbf{X}_{c,i}^{\text{sat}}\bar{\mathbf{f}}}{\Bar{\mathbf{f}}^{\sf H}\mathbf{U}_{c,i}^{\text{sat}}\Bar{\mathbf{f}}}\right)\right)} + \frac{ \left( \mathbf{S}_{p,u}^{\textnormal{sat}} + \mathbf{U}_{p,u}^{\textnormal{sat}} \right) \Bar{\mathbf{f}} }{\Bar{\mathbf{f}}^{\sf H}\left( \mathbf{S}_{p,u}^{\textnormal{sat}} + \mathbf{U}_{p,u}^{\textnormal{sat}} \right) \Bar{\mathbf{f}} \cdot L\left(\Bar{\mathbf{f}}\right)} + \frac{ L\left(\Bar{\mathbf{f}}\right) }{ \nabla_{\Bar{\mathbf{f}}} L\left(\Bar{\mathbf{f}}\right) } \right\}
        + \sum\limits_{k \in \CMcal{K}_t^{{\text{int}}}} \left\{\frac{\exp\left(-\frac{1}{\mu}\log_2 \left( \frac{ \bar{\mathbf{f}}^{\sf H} \mathbf{X}_{c,k}^{\text{bs}} \bar{\mathbf{f}}}{ \bar{\mathbf{f}}^{\sf H} \mathbf{Y}_{c,k}^{\text{bs}} \bar{\mathbf{f}}} \right) \right) \frac{ \mathbf{X}_{c,k}^{\text{bs}} \bar{\mathbf{f}} }{ \bar{\mathbf{f}}^{\sf H} \mathbf{X}_{c,k}^{\text{bs}} \bar{\mathbf{f}}} }{ \sum_{j \in \CMcal{K}_t^{{\text{int}}}} \exp\left(-\frac{1}{\mu}\log_2\left( \frac{ \bar{\mathbf{f}}^{\sf H} \mathbf{X}_{c,j}^{\text{bs}} \bar{\mathbf{f}}}{ \bar{\mathbf{f}}^{\sf H} \mathbf{Y}_{c,j}^{\text{bs}} \bar{\mathbf{f}}} \right) \right)} \right\} \nonumber\\
        &= \sum\limits_{u=1}^{K_s} \left\{ \frac{\exp\left(-\frac{1}{\mu}\log_2\left( \frac{\bar{\mathbf{f}}^{\sf H}\mathbf{X}_{c,u}^{\text{sat}}\bar{\mathbf{f}}}{\bar{\mathbf{f}}^{\sf H}\mathbf{U}_{c,u}^{\text{sat}}\bar{\mathbf{f}}}\right)\right)  \frac{\mathbf{U}_{c,u}^{\text{sat}}\bar{\mathbf{f}}}{\bar{\mathbf{f}}^{\sf H}\mathbf{U}_{c,u}^{\text{sat}}\bar{\mathbf{f}}}}{\sum_{i=1}^{K_s}\exp\left(-\frac{1}{\mu}\log_2\left( \frac{\bar{\mathbf{f}}^{\sf H}\mathbf{X}_{c,i}^{\text{sat}}\bar{\mathbf{f}}}{\Bar{\mathbf{f}}^{\sf H}\mathbf{U}_{c,i}^{\text{sat}}\Bar{\mathbf{f}}}\right)\right)} + \frac{ \mathbf{U}_{p,u}^{\textnormal{sat}} \Bar{\mathbf{f}} }{\Bar{\mathbf{f}}^{\sf H} \mathbf{U}_{p,u}^{\textnormal{sat}} \Bar{\mathbf{f}} \cdot L\left(\Bar{\mathbf{f}}\right) } \right\} + \sum\limits_{k \in \CMcal{K}_t^{{\text{int}}}} \left\{\frac{\exp\left(-\frac{1}{\mu}\log_2 \left( \frac{ \bar{\mathbf{f}}^{\sf H} \mathbf{X}_{c,k}^{\text{bs}} \bar{\mathbf{f}}}{ \bar{\mathbf{f}}^{\sf H} \mathbf{Y}_{c,k}^{\text{bs}} \bar{\mathbf{f}}} \right) \right) \frac{ \mathbf{Y}_{c,k}^{\text{bs}} \bar{\mathbf{f}}}{ \bar{\mathbf{f}}^{\sf H} \mathbf{Y}_{c,k}^{\text{bs}} \bar{\mathbf{f}}} }{ \sum_{j \in \CMcal{K}_t^{{\text{int}}}} \exp\left(-\frac{1}{\mu}\log_2\left( \frac{ \bar{\mathbf{f}}^{\sf H} \mathbf{X}_{c,j}^{\text{bs}} \bar{\mathbf{f}}}{ \bar{\mathbf{f}}^{\sf H} \mathbf{Y}_{c,j}^{\text{bs}} \bar{\mathbf{f}}} \right) \right)} \right\}.\label{KKT_cond}
    \end{align}
    }}
    \hrule
\end{figure*}
we first obtain the partial derivative of $f_1\left(\bar{\mathbf{f}}\right)$ as given by \eqref{L1_deriv} at the top of the next page, where 
\begin{align}
    \mathbf{X}_{c,k}^{\text{bs}} &= \mathbf{S}_{c,k}^{\textnormal{bs}} + \mathbf{C}_{c,k}^{\textnormal{bs}} + \omega_k \mathbf{I}_{M(K_s + 1)},\label{X_bs}\\
    \mathbf{Y}_{c,k}^{\text{bs}} &= \mathbf{C}_{c,k}^{\textnormal{bs}} + \omega_k \mathbf{I}_{M(K_s + 1)},\label{Y_bs}\\
    \mathbf{X}_{c,u}^{\text{sat}} &= \mathbf{S}_{c,u}^{\textnormal{sat}} + \mathbf{U}_{c,u}^{\textnormal{sat}},\label{X_sat}\\
    L\left(\Bar{\mathbf{f}}\right) &= \left\{\prod_{j=1}^{K_t}\Bar{\mathbf{f}}^{\sf H}\left( \mathbf{C}_{p,j}^{\textnormal{bs}} + \epsilon_j \mathbf{I}_{M(K_s+1)} \right)\Bar{\mathbf{f}} \right\}^{\frac{1}{K_s}}.\label{leakage}
\end{align}
Subsequently, we calculate \eqref{deriv_f1}, and rearrange it to satisfy the first-order KKT condition as shown \eqref{KKT_cond} at the top of the next page.
By substituting \eqref{lambda_sat}, \eqref{A}, and \eqref{B} into \eqref{KKT_cond}, it is presented as a eigenvector dependent NEPv as follows.
\begin{align}
    &\mathbf{A}\left(\bar{\mathbf{f}}\right)\bar{\mathbf{f}} = \lambda^{\text{sat}}\left(\bar{\mathbf{f}}\right)\mathbf{B}\left(\bar{\mathbf{f}}\right)\bar{\mathbf{f}} \nonumber\\
    &\Leftrightarrow \mathbf{B}^{-1}\left(\bar{\mathbf{f}}\right)\mathbf{A}\left(\bar{\mathbf{f}}\right)\bar{\mathbf{f}} = \lambda^{\text{sat}}\left(\bar{\mathbf{f}}\right)\bar{\mathbf{f}}.
\end{align}

Second, similarly to the above procedure, the partial derivative of $f_2\left(\bar{\mathbf{v}}\right)$ is given by
\begin{align}
    \frac{\partial f_2\left(\bar{\mathbf{v}}\right)}{\partial \bar{\mathbf{v}}} 
        &= \frac{1}{\log 2} \sum_{j=1}^{K_t} \left[ \frac{\mathbf{S}_{p,k}^{\textnormal{bs}}\bar{\mathbf{v}}}{\bar{\mathbf{v}}^{\sf H}\mathbf{S}_{p,k}^{\textnormal{bs}}\bar{\mathbf{v}}} - \frac{\mathbf{U}_{p,k}^{\textnormal{bs}}\bar{\mathbf{v}}}{\bar{\mathbf{v}}^{\sf H}\mathbf{U}_{p,k}^{\textnormal{bs}}\bar{\mathbf{v}}} \right].\label{KKT_cond_f2}
\end{align}
Using this, after calculating \eqref{deriv_f2}, by substituting equations \eqref{lambda_bs}, \eqref{C}, and \eqref{D} into \eqref{KKT_cond_f2}, it is shown that the first-order KKT condition for $f_2\left(\bar {\bf{v}}\right)$ is satisfied as indicated in \eqref{KKT_bs}.
The proof of $f_2\left(\bar {\bf{v}}\right)$ is omitted since it is straightforward from the proof of $f_1(\bar {\bf{f}})$.
This completes the proof.\qed }}

\bibliographystyle{IEEEtran}
\bibliography{ref_dist_stin.bib}

\end{document}